\documentclass[manuscript,nonacm]{acmart}
\usepackage[linesnumbered,ruled,vlined]{algorithm2e}
\usepackage{threeparttable}
\usepackage{tabularx}
\usepackage{hyperref}
\AtBeginDocument{%
  }

\begin{document}

\title{ChatNeuroSim: An LLM Agent Framework for Automated Compute-in-Memory Accelerator Deployment and Optimization}

\author{Ming-Yen Lee}
\email{mlee838@gatech.edu}
\orcid{0000-0002-5076-4645}
\affiliation{%
  \institution{Georgia Institute of Technology}
  \city{Atlanta}
  \state{Georgia}
  \country{USA}
}

\author{Shimeng Yu}
\email{shimeng.yu@ece.gatech.edu}
\orcid{0000-0002-0068-3652}
\affiliation{%
  \institution{Georgia Institute of Technology}
  \city{Atlanta}
  \state{Georgia}
  \country{USA}
}

\renewcommand{\shortauthors}{M.-Y. Lee and S. Yu}
\thanks{This work is supported in part by PRISM, one of the SRC/DARPA
JUMP 2.0 centers.}
\begin{abstract}
  Compute-in-Memory (CIM) architectures have been widely studied for deep neural network (DNN) acceleration by reducing data transfer overhead between the memory and computing units. In conventional CIM design flows, system-level CIM simulators (such as NeuroSim) are leveraged for design space exploration (DSE) across different hardware configurations and DNN workloads. However, CIM designers need to invest substantial effort in interpreting simulator manuals and understanding complex parameter dependencies. Moreover, extensive design–simulation iterations are often required to identify optimal CIM configurations under hardware constraints. These challenges severely prolong the DSE cycle and hinder rapid CIM deployment. To address these challenges, this work proposes ChatNeuroSim, a large language model (LLM)-based agent framework for automated CIM accelerator deployment and optimization. ChatNeuroSim automates the entire CIM workflow, including task scheduling, request parsing and adjustment, parameter dependency checking, script generation, and simulation execution. It also integrates the proposed CIM optimizer using design space pruning, enabling rapid identification of optimal configurations for different DNN workloads. 
  ChatNeuroSim is evaluated on extensive request-level testbenches and demonstrates correct simulation and optimization behavior, validating its effectiveness in automatic request parsing and task execution. Furthermore, the proposed design space pruning technique accelerates CIM optimization process compared to no-pruning baseline. In the case study optimizing Swin Transformer Tiny under 22 nm technology, the proposed CIM optimizer achieves a 0.42x–0.79x average runtime reduction compared to the same optimization algorithm without design space pruning.
\end{abstract}




\maketitle

\section{Introduction}

\begin{figure}[h]
  \centering
  \includegraphics[width=\linewidth]{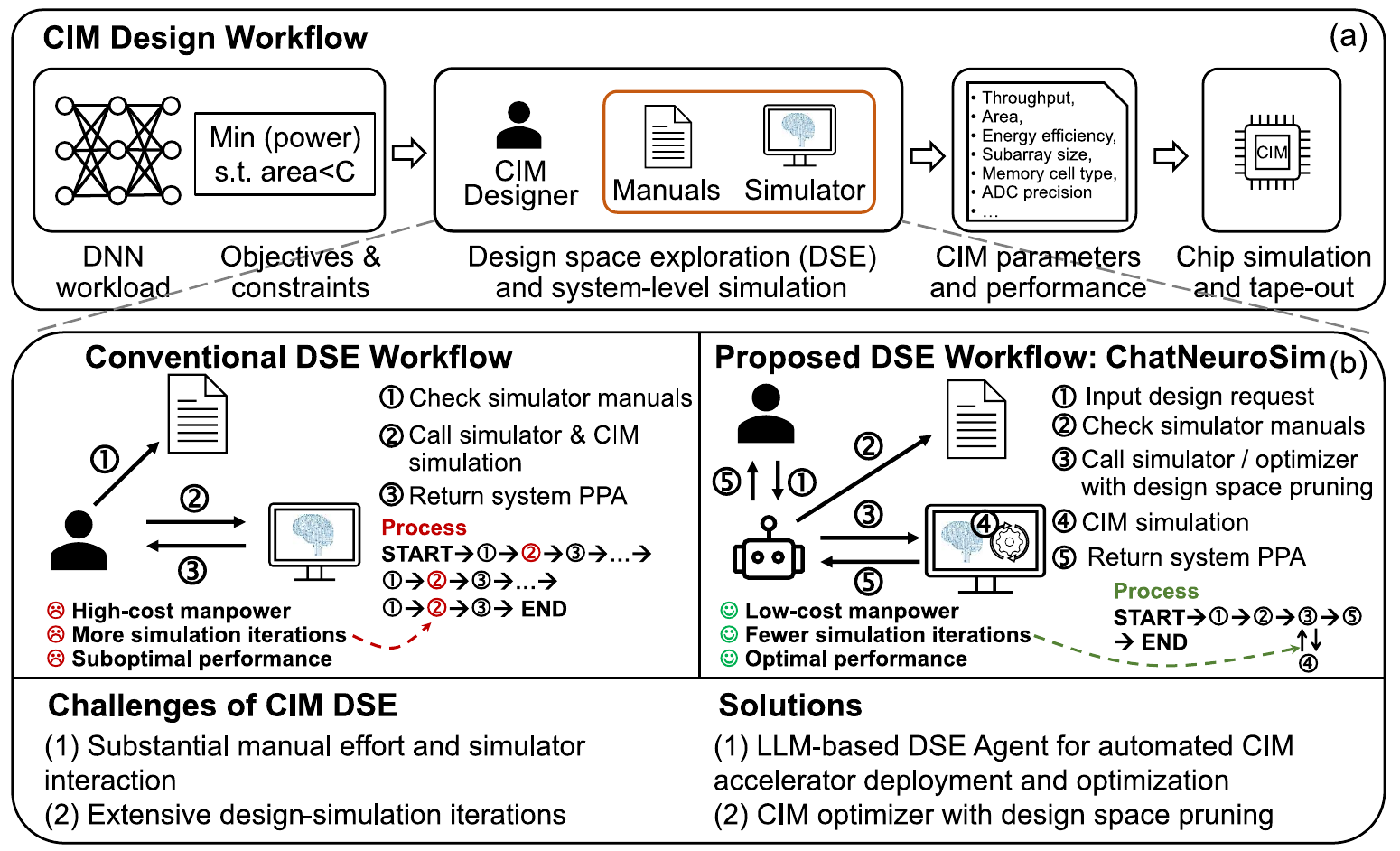}
  \caption{(a) Compute-in-Memory (CIM) design workflow, (b) Conventional design space exploration (DSE) workflow and proposed LLM-based DSE workflow.}
  \label{introduction}
\end{figure}

Specialized hardware accelerators with systolic array, and digital/analog compute-in-memory (CIM) architectures for AI workloads have been increasingly demanded to meet the performance and energy-efficiency requirements of modern AI workloads \cite{intro_lu2024high, intro_sharda2024accelerator, intro_40nmrram}. To reduce time-to-market and design costs,  system-level AI hardware simulators for systolic arrays \cite{intro_scalesim},  \cite{intro_timeloop}, and digital/analog CIM accelerators \cite{intro_neurosim1p5,intro_aihwkit,intro_cimloop} have been adopted to enable efficient design space exploration (DSE) prior to chip tape-out \cite{intro_dse}. 

Despite their effectiveness, these simulators introduce substantial usability challenges. Designers need to invest significant effort to understand simulator functionalities, complex parameter dependencies, and heterogeneous input–output formats by extensively consulting manuals and documentation. Translating high-level design intent into valid simulator parameters, invoking simulations, and parsing output results often require careful manual intervention. As illustrated in Fig. \ref{introduction}, a conventional CIM DSE workflow involves prolonged effort in understanding simulator behavior (Step 1), manual request-to-parameter conversion and simulation execution (Step 2), and post-processing of simulation outputs (Step 3). This iterative process leads to high manpower costs and significantly prolongs the CIM design cycle.

Recently, large language model (LLM)-based multi-agent frameworks have emerged as promising solutions to automate eletronic design automation (EDA) workflows by interpreting documentation, translating user requests into simulator inputs, and managing task execution \cite{intro_chateda,intro_chatarch}. For example, ChatEDA \cite{intro_chateda} integrates LLM-based agents to automate the design flow from register-transfer level (RTL) to graphic data system (GDS), while ChatArch \cite{intro_chatarch} proposes an LLM-driven framework for automated processor architecture optimization. However, despite the availability of multiple CIM simulators \cite{intro_neurosim1p5,intro_aihwkit,intro_cimloop}, the application of LLM-based agent frameworks to automate CIM simulation and deployment remains unexplored.

In addition to the usability challenge, CIM designers face an additional challenge in efficiently exploring the large and complex design space. Even with proficient simulator usage, identifying optimal hardware configurations under multiple resource constraints remains difficult. Designers often rely on extensive domain expertise and repeated simulations to refine candidate configurations, yet the final design might still be suboptimal. To mitigate this challenge, machine learning (ML)-based optimization algorithms such as genetic algorithms, Bayesian optimization, and reinforcement learning (RL), have been introduced to reduce design–simulation iterations and accelerate DSE \cite{intro_apollo,intro_data_driven_google,intro_airchitectv2,intro_kao2020confuciux,intro_kao2020gamma}. For instance, \cite{intro_kao2020confuciux} applies RL and genetic algorithms to balance compute and memory resources for target DNN workloads, while \cite{intro_apollo} presents an RL-assisted architecture exploration framework with transfer learning to reduce the number of required iterations. 

In the field of CIM DSE, prior work \cite{intro_neurosimagent} has proposed an RL-based automated optimization framework that incorporates transfer learning and design space pruning, primarily targeting conventional convolutional neural networks such as ResNet and DenseNet. Nevertheless, several important opportunities remain underexplored, including optimization for more complex workloads such as vision transformers, as well as a systematic evaluation of the generality and effectiveness of design space pruning across different optimization objectives and constraints.

To address these challenges, this work proposes ChatNeuroSim, an LLM-based agent framework on top of NeuroSim for automated CIM accelerator deployment and optimization. ChatNeuroSim automates the CIM DSE workflow by parsing user requests, consulting simulator documentation, generating execution scripts, invoking simulators or optimizers, and returning system-level accuracy, performance, power, and area (PPA) metrics. In addition, we develop a CIM optimizer based on heuristic ML algorithms with design space pruning, extending prior work \cite{intro_neurosimagent} to support rapid optimization across diverse DNN workloads, including ResNet, Swin Transformer \cite{intro_swint}, and Vision Transformer \cite{intro_vit}. The main contributions of this work are summarized as follows:

\begin{itemize}
\item \textbf{An LLM-based agent framework for automated CIM deployment and optimization:}
The proposed framework leverages LLM-based agents to automatically execute the end-to-end CIM workflow, from parsing user requests to invoking CIM simulator (i.e., NeuroSim \cite{intro_neurosim1p5}) or optimizers, and returning final accuracy and PPA results.
\item \textbf{Design space pruning for efficient CIM optimization:}
We propose a heuristic ML-based CIM optimizer with design space pruning, where the exploration space is initialized using knowledge transferred from prior workloads and iteratively refined during optimization. This approach improves search efficiency and reduces optimization runtime compared to conventional heuristic ML baselines.
\item \textbf{Comprehensive agent evaluation and optimization analysis:}
We evaluate the proposed agent framework using a customized CIM request dataset with 40 test cases, achieving 100\% correctness in script generation and simulation results. We also validate the CIM optimizer on Swin Transformer Tiny under a 22 nm technology node using state-of-the-art emerging non-volatile memories. The proposed optimizer with design space pruning achieves 0.42x–0.79x average runtime reduction and 0.29x–0.69x P95 runtime reduction compared to baseline no-pruning algorithms across different optimization objectives. We also analyze the generality of design space pruning under different performance metrics, hardware constraints, and pruning hyper-parameters, and integrate the resulting insights into ChatNeuroSim for user guidance.
\end{itemize}

The remainder of this work is organized in the following order: Section 2 provides background on CIM simulation tools, ML-based accelerator optimization, and LLM-based EDA agents. Section 3 presents the ChatNeuroSim framework, including its agent workflow and optimization process with design space pruning. Section 4 evaluates ChatNeuroSim using diverse CIM requests and benchmark workloads. Section 5 draws the conclusion of this work. The Appendix shows ChatNeuroSim user interface (UI) and dialogue example, with a GitHub link (\url{https://github.com/neurosim/ChatNeuroSim}) for interested users for trial runs.

\section{Background}
\begin{figure}[h]
  \centering
  \includegraphics[width=\linewidth]{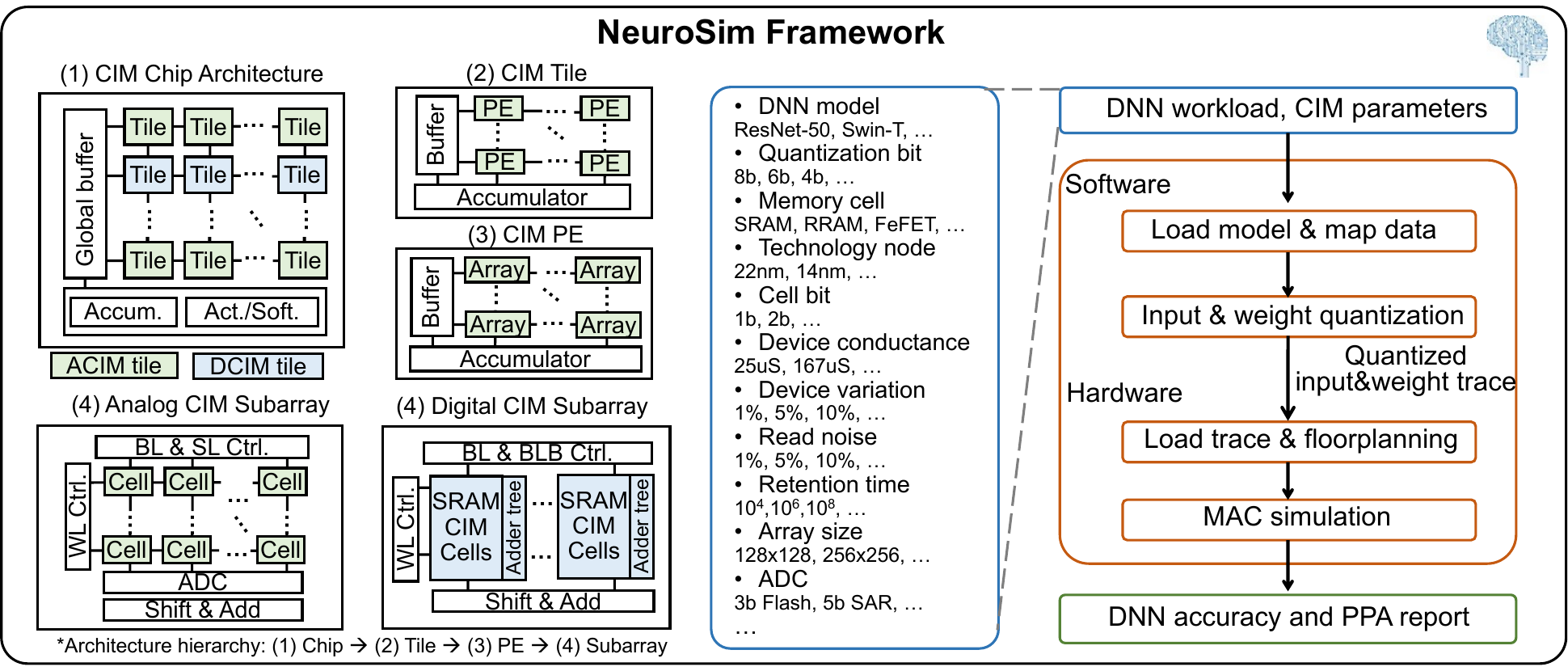}
  \caption{Compute-in-Memory (CIM) simulator: NeuroSim framework.}
  \label{background}
\end{figure}

\subsection{Compute-in-Memory (CIM) and CIM Simulators}

Compute-in-Memory (CIM) is an emerging computing paradigm in which multiply-and-accumulate (MAC) operations of matrix multiplication are directly performed within memory arrays. By eliminating frequent data transfers between memory and compute units, CIM architectures significantly reduce data movement overhead compared to conventional von Neumann architectures. As shown in Fig. \ref{background}, a typical CIM system adopts a hierarchical organization through chip, tile, processing element (PE), and subarray levels. DNN weights are stored in memory cells within subarrays, while input activations are applied to wordlines (WLs). MAC operations are executed intrinsically in the memory array, and the results are either digitized by analog-to-digital converters (ADCs) in analog CIM (ACIM) subarrays or accumulated through digital adder trees in digital CIM (DCIM) subarrays.

Designing a complete CIM system under multiple hardware constraints and performance objectives remains a complex and time-consuming task. To facilitate CIM architecture design and system-level optimization, several CIM simulation frameworks have been proposed \cite{intro_neurosim1p5,intro_aihwkit,intro_cimloop}. By jointly modeling DNN workloads and hardware characteristics, these simulators reduce the design effort and iteration time during design space exploration (DSE) compared to transistor-level simulations.

In these frameworks, NeuroSim \cite{intro_neurosim1p5, intro_dnn+neurosim, intro_lee2024neurosim} has emerged as a comprehensive CIM simulation platform that integrates modeling of DNN workloads, CIM architectures, memory device characteristics and peripheral logic circuits in advanced technology nodes through a unified user-configurable interface. As illustrated in Fig. \ref{background}, after receiving user-defined parameters, NeuroSim first loads the DNN model and architectural configurations. Inputs and weights are quantized and mapped onto CIM subarrays for accuracy simulation with the modeling of hardware non-idealities such as device variations and analog-to-digital conversion (ADC) quantization errors. The resulting quantized data traces are then forwarded to hardware-level models to estimate system-level PPA metrics. The latest NeuroSim V1.5 \cite{intro_neurosim1p5} supports deployment of both convolutional neural networks (CNNs) and vision transformers (ViTs) on CIM architectures for DNN inference. CNN workloads are implemented using ACIM subarrays, while ViTs employ a hybrid ACIM–DCIM architecture, where ACIM is used for linear operations (e.g., Q, K, V projections and MLP layers) and DCIM is used for attention score computation ($\mathrm{QK}^\mathsf{T}$) and weighted aggregation. This joint software–hardware modeling enables end-to-end evaluation of CIM systems and provides critical performance insights prior to silicon tape-out.

\subsection{CIM Design Space Exploration (DSE)}
Although CIM simulators accelerate evaluation compared to transistor-level simulation and optimization, exploring the large and highly coupled CIM design space remains challenging. Based on the design space example in Fig. \ref{background}, the number of tunable hardware parameters can easily lead to a design space size on the order of $10^3$–$10^4$. Efficiently navigating this space requires extensive domain expertise and manual tuning. Meanwhile infeasible configurations that violate area, power, or performance constraints have to be filtered out. Therefore, manual trial-and-tuning approaches become increasingly inefficient as design complexity grows. An automated and efficient CIM optimization methodology is critical to accelerate the DSE process and enable rapid identification of high-performing hardware configurations.

\subsection{Machine Learning (ML) Algorithms for AI Accelerator DSE}
To address the scalability challenges of DSE, a wide range of machine learning (ML) algorithms have been proposed for automated AI accelerator optimization. Simulated annealing \cite{bg_simulated_annealing} is a heuristic search algorithm that probabilistically accepts worse-performing design points based on an acceptance "temperature" to avoid the search process stuck into local optima. As the "temperature" decreases, the exploration gradually focuses on high-potential regions of the design space. Genetic algorithm \cite{intro_kao2020confuciux} is another heuristic search algorithm which represents design points as gene sequences and evolves candidate solutions through mutation and crossover operations across generations. Bayesian optimization constructs a surrogate model to estimate the objective function and identify promising configurations with high expected performance. The Tree-structured Parzen Estimator (TPE) \cite{bg_tpe} is a Bayesian optimization technique that models parameter distributions conditioned on observed performance with efficient sampling of high-quality design points.

Reinforcement learning (RL) provides another effective paradigm by modeling the DSE process as an agent–environment interaction, where actions correspond to selecting hardware parameters and rewards reflect performance objectives. Policy-based RL methods \cite{bg_policy1_sutton1999policy,bg_policy2_schulman2017proximal} are well suited for continuous or high-dimensional parameter spaces, while value-based methods \cite{bg_qbased1_watkins1992qlearning,bg_qbased2_mnih2013playing,bg_qbased3_van2016deep}, such as Q-learning, are particularly effective for discrete design spaces through iterative value estimation. These ML-driven approaches significantly improve the efficiency of AI accelerator DSE by reducing reliance on manual tuning.

\subsection{Large Language Model (LLM)-based Agent Framework for EDA}

Large language models (LLMs) have recently demonstrated extraordinary capabilities in natural language processing applications such as conversational interaction, code synthesis, and document understanding \cite{bg_verilogcoder,intro_chateda,intro_chatarch}. These capabilities have spurred growing interest in applying LLMs to EDA workflows. VerilogCoder \cite{bg_verilogcoder} leverages LLMs for automated Verilog code generation with syntax checking and functional validation. ChatEDA \cite{intro_chateda} introduces an LLM-based agent interface that automates the circuit design flow from register-transfer level (RTL) generation to graphic data system II (GDSII) implementation. By decomposing user requests into structured sub-tasks, these multi-agent frameworks automate task scheduling, script generation, syntax verification, and tool invocation with minimal human intervention.

Prompt engineering \cite{bg_prompt_engineering} and Retrieval-Augmented Generation (RAG) \cite{bg_rag} are two key techniques for constructing effective LLM-based multi-agent systems without requiring model retraining or fine-tuning. Prompt engineering provides structured input–output templates, reasoning guidelines, and execution constraints that guide LLM behavior during task processing. Carefully designed prompts lead to consistent parsing of user intent, reliable parameter extraction, and deterministic tool invocation. RAG  further enhances reliability by retrieving relevant domain knowledge (e.g., simulator manuals and parameter specifications) from an external knowledge base before response generation. By grounding LLM reasoning in retrieved authoritative information, RAG significantly improves response accuracy and reduces hallucination during complex EDA tasks.

Given the benefits of LLM-based multi-agent systems in EDA domain, developing a dedicated LLM-driven agent framework for CIM is of great interest to the research community. Such a framework can reduce human effort, improve design productivity, and accelerate CIM deployment.

\section{ChatNeuroSim: LLM-based agent framework of automated CIM deployment and optimization}

\subsection{Overall Framework}
\begin{figure}[h]
  \centering
  \includegraphics[width=\linewidth]{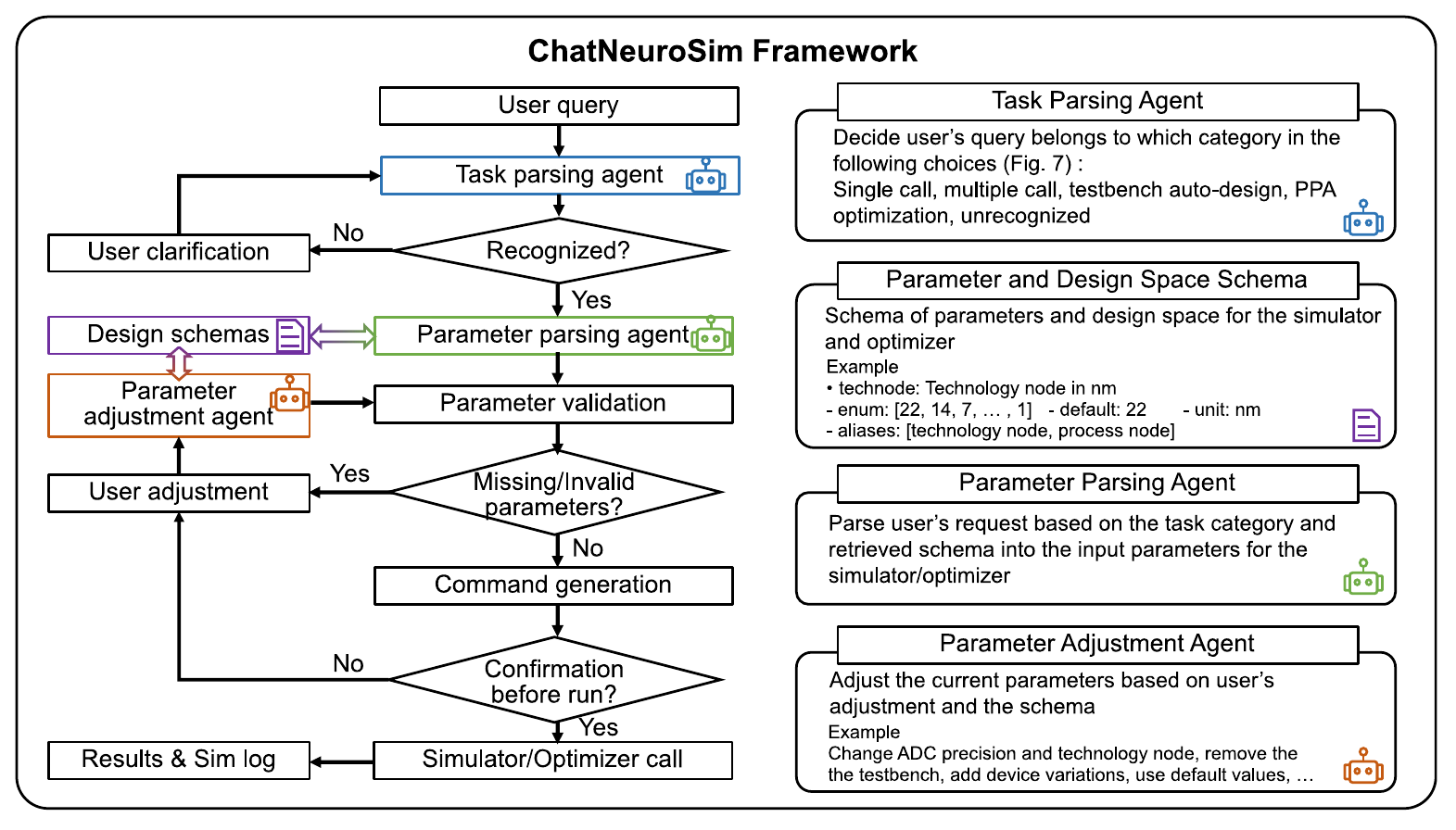}
  \caption{ChatNeuroSim: Overall framework.}
  \label{chatNSFramework}
\end{figure}

\begin{table}[h]
\centering
\begin{threeparttable}
\caption{CIM Request Categories}
\label{tab:requestCategory}{
\renewcommand{\arraystretch}{1.3}
\begin{tabularx}{\linewidth}{
>{\hsize=0.5\hsize\arraybackslash}X
>{\hsize=0.8\hsize\arraybackslash}X
>{\hsize=1.7\hsize\arraybackslash}X}
\toprule
\textbf{Query Category} & \textbf{Description} & \textbf{Example} \\ 
\midrule
Single Call & Single testbench and one simulation with one fixed set of parameters. & I want to simulate VGG8 on CIFAR-10 on CIM architecture using 8b quantization for both input and weight. I only want to get the PPA estimation. The memory device is 22nm RRAM with 1bit cell precision. The state conductances are 25uS and 167uS with no variations. The subarray size is 128x128 and ADC precision 7bit. \\ 
Multiple Call & Multiple testbenches and simulations with different user’s specified parameter values. & I want to simulate VGG8 on CIFAR-10 on CIM architecture using 8b quantization for both input and weight. I only want to get the PPA estimation under 22nm, 14nm, and 7nm tech node. The memory device is SRAM. The subarray size is 128x128 and ADC precision 7bit. \\ 
Testbench \newline Auto-design & Multiple testbenches and simulations with user’s  specified parameter names but without specified parameter values. & Check the accuracy drop of ResNet-18 on CIFAR-100 of different input and weight quantization precisions. Only simulate the accuracy without the PPA. Run 5 batches with the batch size 100. Use 1b cell with a subarray of 128x128 and 7b ADC. \\ 
PPA Optimization & PPA optimization request to find the best CIM configuration under constraints. & I’d like to design an CIM chip within 3600mm$^2$ based on 22nm tech node running ResNet-50 on ImageNet with the minimum power consumption. Use simulate annealing to find out the optimal parameter combination of device type, ADC type, ADC precision, subarray size, subarray mux with 20 episodes. \\ 
\bottomrule
\end{tabularx}
}
\end{threeparttable}
\end{table}

Fig. \ref{chatNSFramework} illustrates the overall framework and workflow of ChatNeuroSim, which consists of three LLM-based agents and a set of schemas for parameter and design space retrieval. ChatNeuroSim accepts a user’s CIM query as the initial input (e.g., Fig. \ref{dialogueExample}) and first invokes a task parsing agent to classify the request into an execution category. 

ChatNeuroSim supports four typical request categories, which are single call, multiple call, testbench auto-design, and PPA optimization. An exceptional case is also integrated when the request cannot be classified. The definitions and representative query examples for these categories are summarized in Table \ref{tab:requestCategory}. A single-call request evaluates the accuracy and/or PPA metrics for a specific hardware configuration. A multiple-call request specifies multiple explicit parameter sets, resulting in multiple simulator calls across different testbenches. The testbench auto-design category also involves multiple simulator calls, but differs in that parameter values are not explicitly specified. Instead, the user expresses a high-level intent (e.g, analyzing PPA trends across SRAM technology scaling) and the parameter parsing agent automatically determines appropriate sweeping values based on the design space schema. The final category, PPA optimization, targets identification of optimal hardware configurations under user-defined constraints. In this case, ChatNeuroSim calls the CIM optimizer (Section \ref{section_optimizer}), which internally interacts with the CIM simulator after parameter parsing and validation. If the task parsing agent cannot confidently determine the request category, it prompts the user for clarification and resubmits an adjusted request.

Once the request category is determined, the parameter parsing agent retrieves the corresponding schema for simulator or optimizer inputs. Each schema entry specifies parameter attributes such as supported values, default values, units, technology nodes, and aliases. Based on these definitions, parameters are extracted and converted from the user’s natural-language query into structured inputs. Category-specific system prompts are designed to ensure that parameters are parsed consistently and formatted correctly for downstream simulation or optimization tasks.

User requests may omit required parameters or include invalid values. To address this, ChatNeuroSim validates all extracted parameters against the schema and provides feedback to the user when missing or invalid entries are detected. The user can then send an adjustment request, which is processed by the parameter adjustment agent to insert missing parameters or modify existing values in accordance with the schema and user intent. After all parameters pass validity checks, ChatNeuroSim generates the execution script and requests user confirmation before execution. Upon confirmation, the framework invokes the simulator or optimizer and returns the PPA metrics and execution logs. 

Throughout the workflow, users interact with ChatNeuroSim using standard CIM terminology without manually inspecting simulator documentation or parameter dependencies. The framework automatically maps high-level user requests to valid simulator or optimizer inputs, supports dynamic adjustment of multiple parameters and simulations, and flexibly adds or removes testbenches. By automating the entire pipeline from user query to simulation results, ChatNeuroSim significantly reduces the time and manual effort required to design and optimize high-performance CIM systems.

\subsection{Prompt Design for Agents}
\begin{figure}[h]
  \centering
  \includegraphics[width=\linewidth]{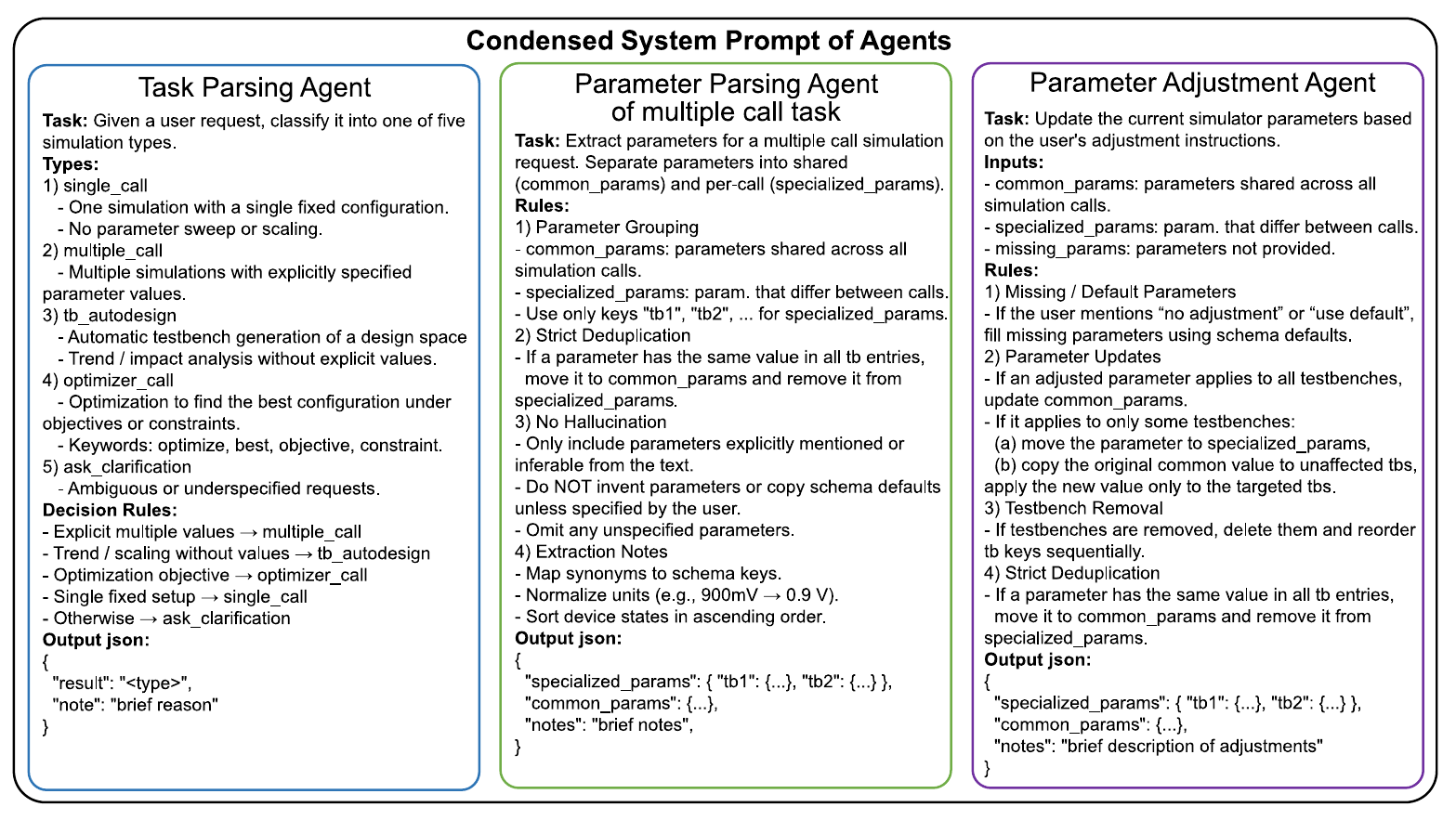}
  \caption{ChatNeuroSim: Prompt design for agents.}
  \label{agentPrompt}
\end{figure}

Each agent in ChatNeuroSim is designed with a dedicated system prompt for a correct, consistent, and reliable behavior in response to user inputs. Fig. \ref{agentPrompt} presents condensed versions of the system prompts for the three agents in ChatNeuroSim, highlighting the essential components required for effective agent operation. Each condensed prompt consists of three key elements: task description, rule specification, and output structure template. Below summarizes the prompt design flow of three agents.

\begin{itemize}
    \item \textbf{Task Parsing Agent:}
    The task parsing agent classifies user requests into predefined execution categories. As summarized in Table \ref{tab:requestCategory}, each category is defined by a distinct set of semantic features. These definitions are explicitly specified in the \textbf{Types} part of the prompt, while the distinguishing characteristics are encoded in the \textbf{Decision Rules}. Based on these structured descriptions, the agent determines the most appropriate request category. To enhance robustness, the prompt also defines an exceptional handling mechanism: if the request cannot be confidently matched to any of the predefined categories, the agent requests additional clarification from the user. The agent outputs both the classification result and the reasons for its decision.

    \item \textbf{Parameter Parsing Agent:}  
    After the request category is determined, the parameter parsing agent retrieves a category-specific prompt to extract structured parameters from the user query. This type-aware prompt design ensures the agent to accommodate heterogeneous parameter formats and schema definitions across different tasks. For example, in a multiple-call task, the agent extracts simulator input parameters defined in the NeuroSim schema, whereas in a PPA optimization task, it retrieves optimizer-specific parameters such as the design space and optimization algorithm configurations.
    
    Each prompt is composed of three components: a task description, parameter extraction rules, and a structured output template. As illustrated in Fig. \ref{agentPrompt}, extracted parameters are organized into two groups: \emph{specialized parameters} (\texttt{specialized\_params}), which vary across testbenches, and \emph{common parameters} (\texttt{common\_params}), which are shared across all testbenches. The agent follows a set of explicit rules, including separation of specialized and common parameters, strict no-hallucination constraints, and format normalization requirements. The output includes both the parsed parameters and key notes during the extraction.

    \item \textbf{Parameter Adjustment Agent:}  
    The parameter adjustment agent processes user requests to modify, add, or remove simulation parameters. For missing parameters, the agent assigns values based on explicit user input or schema-defined defaults. A key principle in this stage is the correct management of specialized and common parameters, particularly in scenarios involving multiple testbenches. Parameters may be switched from specialized to common scope or transferred in the opposite direction after the adjustment. In addition, when testbenches are added or removed, the agent needs to reorder testbench indices to maintain consistency. These rules are encoded directly into the system prompt for reliable parameter updates. The agent outputs the revised parameter sets with detailed adjustment notes.

\end{itemize}

\begin{figure}[h]
  \centering
  \includegraphics[width=\linewidth]{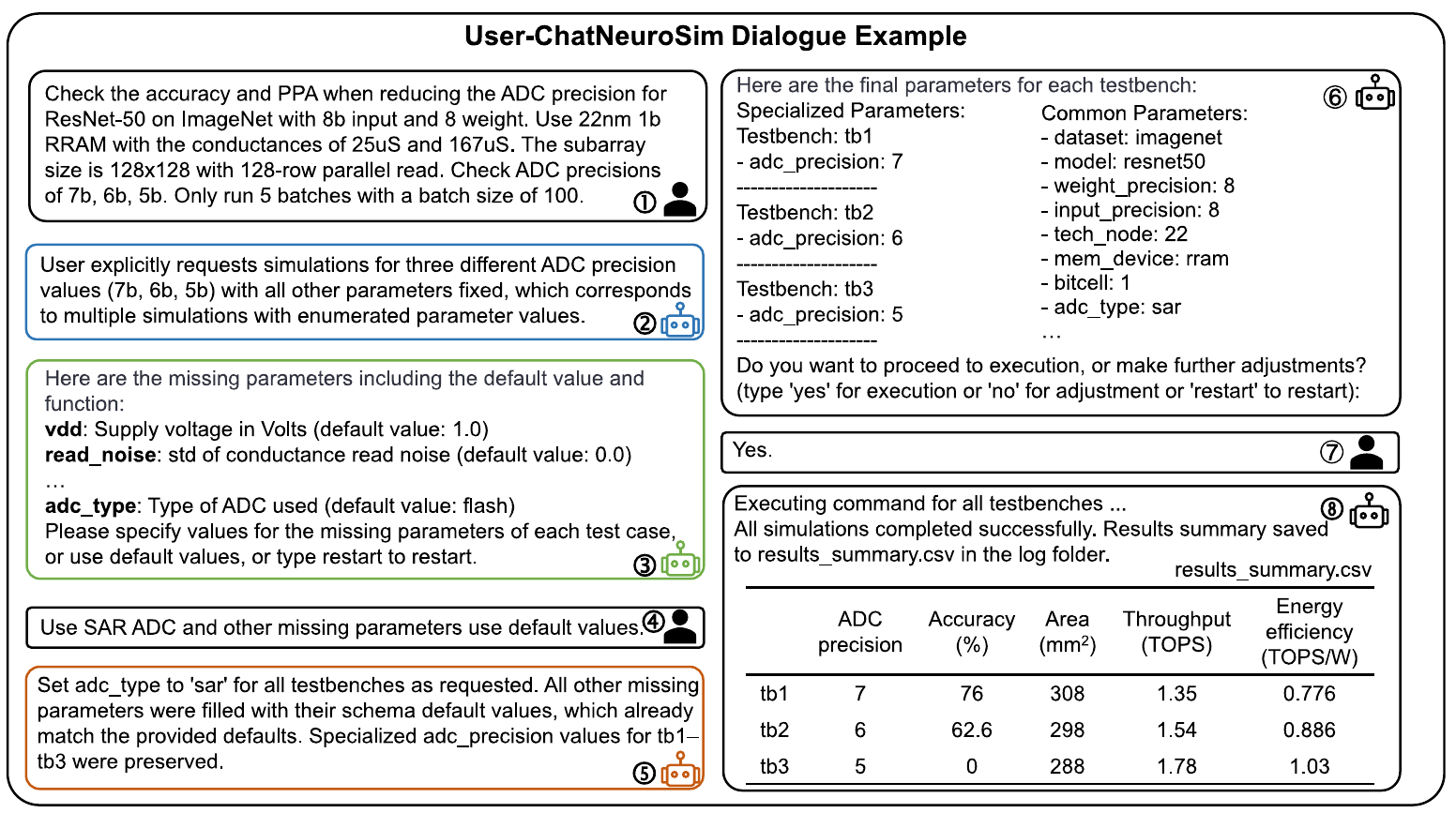}
  \caption{Dialogue example between the user and ChatNeuroSim.}
  \label{dialogueExample}
\end{figure}

Fig. \ref{dialogueExample} shows a representative dialogue between the user and ChatNeuroSim, demonstrating the end-to-end interaction across request submission, parameter adjustment, agent responses, and final simulation execution. In this example, the user issues a multiple-call request to evaluate accuracy and PPA for ResNet-50 using three different ADC precisions. The task parsing agent correctly identifies the request category based on the prompt-defined criteria. The parameter parsing agent then extracts simulation parameters and reports missing entries required for execution. After the user supplies the missing values such as default parameters and ADC precision, the parameter adjustment agent updates the parameter sets accordingly. The finalized configurations for the three testbenches are organized into specialized and common parameter groups, as shown in Step 6 of Fig. \ref{dialogueExample}.

As the user confirms the simulation execution, ChatNeuroSim generates the execution script, invokes NeuroSim, and records the simulation results and logs. Through the interaction, users only input standard CIM terminology without manually consulting simulator documentation or constructing execution commands. By automatically mapping high-level user intent to valid parameter configurations and executable scripts, ChatNeuroSim performs seamless simulation and result inspection. This automated, agent-driven workflow substantially reduces design turnaround time and alleviates manual effort in CIM system exploration and optimization.

\subsection{CIM Optimizer with Design Space Pruning} \label{section_optimizer} 

\begin{figure}[h]
  \centering
  \includegraphics[width=\linewidth]{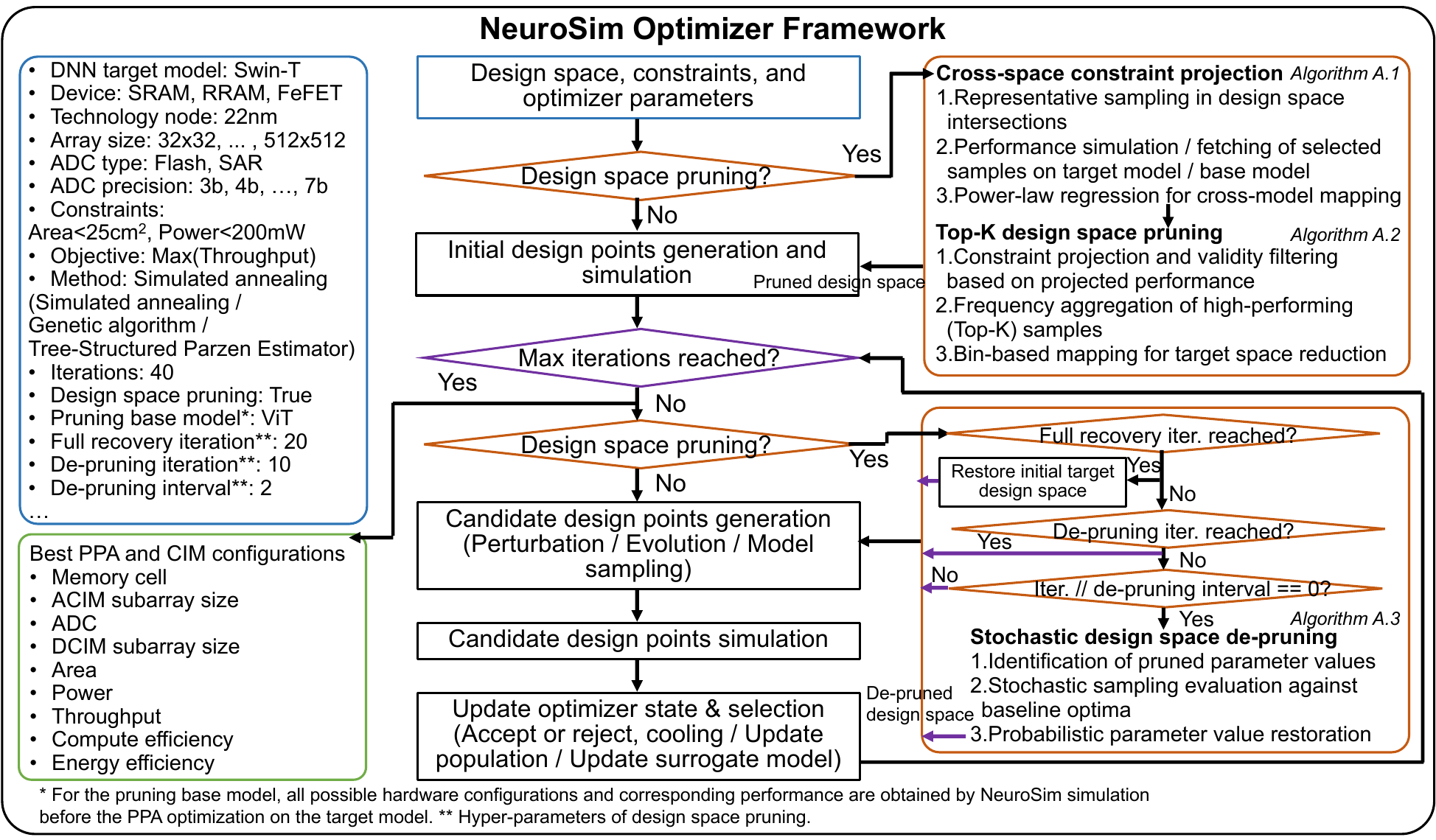}
  \caption{ChatNeuroSim: CIM optimizer framework.}
  \label{optimizerFramework}
\end{figure}

In addition to the agent framework, ChatNeuroSim incorporates a CIM optimizer supporting automated PPA optimization through efficient design space search. As shown in Fig. \ref{optimizerFramework}, the optimizer accepts (1) the optimization objective for a target DNN workload, (2) a hardware design space specification (e.g., memory device type, subarray dimensions, ADC configuration), (3) hardware constraints such as power and area, (4) the selected heuristic optimization algorithm, and (5) hyper-parameters controlling the proposed design space pruning algorithms. During optimization, the optimizer repeatedly interacts with NeuroSim by simulating candidate CIM configurations, and finally reports the best PPA and corresponding hardware configuration that satisfies the constraints.

\noindent\textbf{Baseline optimization without design space pruning}

\noindent After receiving user inputs, the optimizer initializes design points and iteratively refines candidate configurations. In the baseline mode, the user selects a standard heuristic algorithm (i.e., simulated annealing (SA), genetic algorithm (GA), or tree-structured Parzen estimator (TPE)) without design space pruning. Although their sampling strategies differ, each iteration can be summarized into three common steps.

\begin{itemize}
    \item \textbf{Candidate generation:}
    SA generates a new candidate by randomly perturbing the current high-performing configuration. GA constructs a new generation by applying mutation and crossover to a set of high-performing “parent” configurations. TPE samples candidates based on a learned surrogate distribution, built from the empirical parameter statistics of previously observed “good” versus “bad” samples.
    \item \textbf{Simulation and record:}
    For each candidate, the optimizer calls the CIM simulator to obtain its PPA, then records the results in a history buffer.
    \item \textbf{State update:}
    The optimizer updates its internal state and the set of best-performing configurations based on the updated history buffer. For example, SA accepts or rejects a candidate according to its objective improvement and a temperature-controlled probability, GA updates the parent pool for the next generation, and TPE updates the parameter distributions used by its surrogate model.
\end{itemize}

After reaching the maximum number of iterations, the optimizer returns the optimal configuration and its PPA.

\noindent \textbf{Motivation for design space pruning}

\noindent While these heuristic algorithms provide effective search capabilities, optimization runtime can remain high when the design space is large and each PPA evaluation is expensive. Particularly, The complex transformer workloads often include larger design space compared to CNN (e.g., ResNet-50) optimization that mainly explores ACIM parameters. Optimizing vision transformers such as ViT-Base (ViT-B) and Swin Transformer Tiny (Swin-T) typically expands the search space by introducing DCIM-related configurations and increases NeuroSim simulation latency (Section \ref{sec:casestudy2}). To further improve search efficiency, we propose a transfer-driven design space pruning strategy that leverages knowledge from prior optimizations on other workloads (referred to as \emph{base models}) to accelerate optimization on a \emph{target model}.

\noindent \textbf{Design space pruning with cross-space constraint projection}

\noindent Fig. \ref{optimizerFramework} outlines the three pruning-related algorithms integrated in the CIM optimizer, with detailed pseudocode provided in Appendix \ref{appendixDSAlgorithms}. The first two procedures (i.e., cross-space constraint projection and Top-$K$ pruning) are executed before the first optimization iteration.

\noindent \textbf{Cross-space constraint projection} (Algorithm \ref{alg:constraint-projection}): When transferring knowledge from a base model to a target model, a key challenge is that feasibility under hardware constraints (e.g., power and area) typically cannot be directly comparable across design spaces. Therefore, we propose a cross-space projection method to map constraint values from the base space to the target space. Motivated by power-law scaling commonly observed in hardware performance trends, we model the relationship between a base-space metric $X$ and its corresponding target-space metric $Y$ as:
\begin{equation}
Y = \alpha X^{\beta},
\end{equation}
where $\alpha$ and $\beta$ capture space-specific scaling effects. By applying $log()$ function, the projection becomes a linear regression in the log-log domain:
\begin{equation}
\ln\left(Y^{m,c}\right) = a_1^{m,c}\ln\left(X^{m,c}\right) + a_0^{m,c},
\end{equation}
where $X^{m,c}$ and $Y^{m,c}$ denote the value of constraint $c$ for memory type $m$ in the base and target design spaces, respectively. The coefficients $(a_0^{m,c}, a_1^{m,c})$ are estimated via ordinary least squares (OLS), as summarized in Algorithm \ref{alg:constraint-projection}.

Algorithm \ref{alg:constraint-projection} proceeds as follows. First, it identifies the intersection design space $\mathcal{I}$ between the base and target spaces with consistent parameters. For each memory type, we select boundary values $\mathcal{V}$ of critical parameters (e.g., subarray dimensions, ADC resolution, mux configurations) and randomly sample candidate points from combinations of these boundaries to construct a fitting set. These candidates are simulated in the target space to obtain $Y^{m,c}$. For the base space, we assume that PPA values have been simulated and stored from prior optimization runs, allowing $X^{m,c}$ to be obtained through a lookup table rather than re-simulation. With paired samples $(X_i^{m,c}, Y_i^{m,c})$, we perform OLS fitting to derive the projection coefficients.

\noindent \textbf{Top-K design space pruning} (Algorithm \ref{alg:topk-pruning}): Using the projection functions from the previous algorithm, Algorithm \ref{alg:topk-pruning} prunes the target design space before the optimization iterations. We first project constraint values for design points in the intersection space from base to target and filter out configurations predicted to violate target constraints. This process avoids running simulations in the target design space. We then compute the value-frequency statistics of each parameter among the Top-K performing samples of the base model in the intersection design space. For the pruning robustness when handling discrete value granularity, parameter values are grouped into \emph{small}, \emph{mid}, and \emph{large} bins. Frequencies are aggregated per bin, and bins exceeding a predefined threshold are retained which construct the high-performing regions. This process yields a pruned target design space that prioritizes values associated with high-performing configurations of the projected base design space.

\noindent \textbf{Stochastic de-pruning for recovery and robustness} (Algorithm \ref{alg:depruning}): 
Since pruning trades exploration for efficiency, the global optimal configuration may not exist in the initial pruned space. To mitigate this risk, we introduce a stochastic de-pruning mechanism (Algorithm \ref{alg:depruning}) that gradually restores excluded values during optimization, balancing exploitation of the pruned space with controlled re-expansion toward the full design space.

Algorithm \ref{alg:depruning} is governed by a de-pruning interval and a full-recovery threshold. At each de-pruning event (every de-pruning interval), the algorithm identifies parameter values removed in the previous pruning and evaluates them through stochastic sampling. Specifically, it replaces the high-performing baseline configurations by the missing value to form candidate designs, compares their performance to the selected baseline, and computes a win-rate for each candidate value. The win-rate is then used as a probabilistic score to decide whether that value should be restored into the current search space. This process repeats periodically until the optimization reaches full-recovery threshold, after which the full original design space is restored for a comprehensive search.

In summary, cross-space constraint projection, Top-K pruning, and stochastic de-pruning provide an efficient and robust CIM optimization for complex DNN workloads. This integrated approach reduces simulation overhead in early-stage exploration while preserving the ability to recover excluded regions and approach globally optimal PPA solutions. A detailed analysis of when these techniques provide the largest benefit under different objectives, constraints, and pruning hyper-parameters is presented in Section \ref{sec:evaluation}.

\section{Evaluations and Case Studies} \label{sec:evaluation}
\subsection{Experimental Setup}\label{sec:setup}

ChatNeuroSim is implemented in Python, and the agent workflow is developed using LangGraph \cite{langgraph2024}. To evaluate correctness and robustness across diverse user intents, we construct a request-level benchmark containing 40 CIM queries, with 10 requests for each task category listed in Table \ref{tab:requestCategory}. For the agent back end API, we use GPT-5.1 \cite{exp_openai2025gpt51}, GPT-5-mini \cite{exp_openai2025gpt5mini}, DeepSeekV3.2-Chat, and DeepSeekV3.2-Reasoner \cite{exp_deepseek2025v32} to validate agent functionality. All ChatNeuroSim agent executions run on an Intel(R) Xeon(R) Gold 5317 CPU. NeuroSim V1.5 simulations are executed on an NVIDIA RTX A6000 GPU and an Intel(R) Xeon(R) Gold 5317 CPU. We also implement a Chatbot-like UI to facilitate interactive usage based on Streamlit Python library \cite{streamlit}. A UI dialogue example and corresponding simulation outputs are provided in Appendix \ref{appendixExample}.

\begin{table}[h]
\centering
\begin{threeparttable}
\caption{CIM Optimizer: Design Space Exploration Parameters}
\label{tab:dseParams}{
\renewcommand{\arraystretch}{1.3}
\begin{tabularx}{10cm}{
>{\hsize=0.5\hsize\arraybackslash}X
>{\hsize=0.5\hsize\arraybackslash}X
>{\hsize=0.5\hsize\arraybackslash}X
>{\hsize=0.5\hsize\arraybackslash}X
>{\hsize=0.5\hsize\arraybackslash}X
>{\hsize=0.5\hsize\arraybackslash}X}
\toprule
\textbf{Parameters} & \textbf{Range / Values} \\ 
\midrule
Model & ResNet-50, Swin-T, ViT-B \\ 
Dataset & ImageNet \\ 
Input / Weight precision & 8b / 8b \\
Technology node & 22nm \\
Device & SRAM, RRAM, FeFET \\
ACIM subarray column / row$^a$ & 32, 64, 128, 256, 512 \\
ADC type & Flash, SAR \\
ADC precision$^b$ & 3b - 7b \\
Columns per ADC & 4, 8, 16, 32 \\
Weight duplication$^c$ & 0, 1 \\
DCIM subarray column / row$^a$ & 32, 64, 128, 256 \\
\bottomrule
\end{tabularx}
}
\begin{tablenotes}
\scriptsize
    \item $a$. ResNet-50 is implemented with only ACIM, Swin-T and ViT-B are implemented with ACIM for linear operations and DCIM for attention score calculation.
    \item $b$. Parallel read equals to $2^{\text{ADC}_\text{precision}}$. ACIM subarray row smaller than $2^{\text{ADC}_\text{precision}}$ is regarded as invalid design point.
    \item $c$. Weight duplication is only applied to ResNet-50’s convolution operations. 0 means no weight duplication for pipeline (long bubble, small area), 1 means weight duplication for pipeline (short bubble, large area).
    \item $d$. Total design space: ResNet-50: 5280, Swin-T: 42240, ViT-B: 42240.
\end{tablenotes}
\end{threeparttable}
\end{table}

For the CIM optimizer evaluation, we test optimization on both CNN and vision transformer workloads under different design spaces, optimization algorithms, hardware constraints, and performance objectives. We further compare runtime and optimized outcomes with and without the proposed design space pruning, and summarize practical insights on when pruning is most beneficial in the subsequent case studies. These insights are integrated into ChatNeuroSim as suggestions during optimization requests.

Table \ref{tab:dseParams} summarizes the DNN workloads and CIM parameters used in the optimization benchmarks. We target inference PPA for three ImageNet models: ResNet-50 \cite{exp_resnet}, Swin Transformer Tiny (Swin-T) \cite{intro_swint}, and Vision Transformer Base (ViT-B) \cite{intro_vit}. Since ResNet-50 has a smaller design space and shorter NeuroSim runtime than Swin-T and ViT-B, the case studies primarily focus on Swin-T as a representative vision transformer workload. Both input and weight precisions are 8-bit in bit-serial compute mode. One operation is defined as one 8-bit addition or one 8-bit multiplication. We evaluate three 22 nm memory technologies: static random access memory (SRAM), resistive random access memory (RRAM) \cite{exp_rram22nm}, and ferroelectric field effect transistor (FeFET) \cite{exp_dunkel2017fefet22nm}. ADC options include Flash and Successive Approximation Register (SAR) architectures with 3–7 bit precision and 4–32 column-sharing settings. To explore performance–area trade-offs, the design space further includes weight duplication options, which can reduce pipeline bubble overhead during layer pipelining \cite{intro_dnn+neurosim}.

Runtime and best-achieved performance are both critical for practical optimization. For runtime evaluation, the CIM optimizer and NeuroSim PPA simulations are executed on an Intel(R) Xeon(R) Gold 5317 CPU with 48 cores. We allocate 32 physical cores to parallel simulation and reserve the remaining cores for operating-system overhead, I/O, and simulator-internal threading.

To identify an appropriate batch size in optimization iterations, the impact of batch size (16, 32, and 48) on search efficiency and runtime is explored. As shown in Table \ref{tab:batchSize}, the number of samples required to reach the target optimal PPA is relatively consistent across batch sizes. Results are computed over 50 independent runs per algorithm. Under 32-core parallelism, the average per-iteration simulation time increases approximately linearly as batch size scales from 16 to 48. Among the evaluated configurations for figure-of-merit (FoM) optimization on Swin-T, batch size 32 yields the lowest average runtime considering three heuristic algorithms. Therefore, in the following optimization case studies, batch size 32 with 32-core parallelism is deployed.

\begin{table}[h]
\centering
\begin{threeparttable}
\caption{Average Simulation Samples of Swin-T FoM Optimization using Different Strategies and Batch Size}
\label{tab:batchSize}{
\renewcommand{\arraystretch}{1.3}
\begin{tabularx}{\linewidth}{
>{\hsize=0.5\hsize\arraybackslash}X
>{\hsize=1.1\hsize\arraybackslash}X
>{\hsize=1.1\hsize\arraybackslash}X
>{\hsize=1.1\hsize\arraybackslash}X
>{\hsize=1.1\hsize\arraybackslash}X
>{\hsize=1.1\hsize\arraybackslash}X}
\toprule
\textbf{Batch Size} & \textbf{GA \newline (Samples)} & \textbf{SA \newline (Samples)} & \textbf{TPE \newline (Samples)} & \textbf{Average Batch \newline Runtime (min)*} & \textbf{Average Total \newline Runtime (min)**}\\ 
\midrule
16 & 1277 & 719 & 1862 & 5.9 & 472 \\ 
32 & 1048 & 622 & 1660 & 7.2 & \textbf{250} \\ 
48 & 1459 & 598 & 1831 & 9.9 & 266 \\
\bottomrule
\end{tabularx}
}
\begin{tablenotes}
\scriptsize
    \item GA: Genetic Algorithm, SA: Simulated Annealing, TPE: Tree-Structured Parzen Estimator.
    \item FoM = Energy efficiency x Compute efficiency, unit: TOPS$^2$/W/mm$^2$, Area Constraint: 25 cm$^2$, Power Constraint: 200 mW
    \item * Average batch runtime is characterized over 30 independent simulation trials with 32-core parallelism.
    \item ** Calculated as $\frac{1}{3} \sum_{alg \in \{GA, SA, TPE\}} (\frac{S_{alg}}{n} \times T_{batch})$, where $S_{alg}$ is the average simulation samples, $n$ is the batch size, and $T_{batch}$ is the average batch runtime.
\end{tablenotes}
\end{threeparttable}
\end{table}

To provide a realistic estimation of DSE runtime, we adopt a characterization-based cost model. Since NeuroSim evaluations require around 7 minutes runtime overhead per batch size, we first characterize the simulator runtime by running 30 independent trials for batch sizes $n \in {1, 8, 16, 24, 32}$. In addition to $n=32$, we include smaller batch sizes because the proposed pruning algorithms introduce verification samples during de-pruning, and some candidates may already be cached simulation results in previous iterations, resulting in a simulation batch size below 32.

During optimization, we record the number of unique design points evaluated in each iteration $i$, denoted $n_i$. Total runtime is estimated as:
\begin{equation}
\mathcal{T}{total} = \sum_{i=1}^{M} \left(T_{logic,i} + T_{batch}(n_i)\right),
\end{equation}
where $M$ is the number of optimization iterations, $T_{logic,i}$ captures the optimizer’s computational overhead (negligible compared to simulation time), and $T_{batch}(n_i)$ is the measured NeuroSim runtime for evaluating $n_i$ design points. For values of $n_i$ not explicitly characterized, we apply linear interpolation within the corresponding interval. Using this model, we report the mean and $P95$ runtime over 50 independent optimization runs, capturing both algorithmic variability and hardware-constrained simulation costs. Unless otherwise specified, all runtime and performance results in the following experiments are based on 50 independent runs.

\subsection{Case Study 1: Accuracy of Request Dataset on ChatNeuroSim}
The first case study evaluates the correctness of ChatNeuroSim in processing different user requests. The evaluation dataset consists of 40 CIM queries, with 10 requests for each task category defined in Table \ref{tab:requestCategory}. A request is considered correct if and only if ChatNeuroSim generates a valid and runnable execution script that completes successfully and produces simulation results. In this study, we exclude the parameter adjustment stage and focus on assessing ChatNeuroSim’s one-shot script generation capability. We evaluate request processing accuracy using four LLM API back ends: GPT-5.1 \cite{exp_openai2025gpt51}, GPT-5-mini \cite{exp_openai2025gpt5mini}, DeepSeekV3.2-Chat, and DeepSeekV3.2-Reasoner \cite{exp_deepseek2025v32}. Table \ref{tab:acc} summarizes the overall accuracy and the category-level breakdown for each model.

\begin{table}[h]
\centering
\begin{threeparttable}
\caption{ChatNeuroSim: Accuracy of Proposed Request Dataset}
\label{tab:acc}{
\renewcommand{\arraystretch}{1.3}
\begin{tabularx}{\linewidth}{
>{\hsize=1.5\hsize\arraybackslash}X
>{\hsize=0.9\hsize\arraybackslash}X
>{\hsize=0.9\hsize\arraybackslash}X
>{\hsize=0.9\hsize\arraybackslash}X
>{\hsize=0.9\hsize\arraybackslash}X
>{\hsize=0.9\hsize\arraybackslash}X}
\toprule
\textbf{LLM / Accuracy (\%)} & \textbf{Single Call} & \textbf{Multiple Call} & \textbf{Testbench \newline Auto-design} & \textbf{PPA \newline Optimization} & \textbf{Total \newline Accuracy}\\ 
\midrule
GPT-5.1 & 100 & 100 & 100 & 100 & 100 \\ 
GPT-5 mini & 100 & 90 & 80 & 90 & 90 \\ 
DeepSeekV3.2-Reasoner & 90 & 100 & 100 & 90 & 95 \\
DeepSeekV3.2-Chat & 80 & 90 & 60 & 80 & 77.5 \\
\bottomrule
\end{tabularx}
}
\end{threeparttable}
\end{table}

As shown in Table \ref{tab:acc}, ChatNeuroSim achieves 100\% accuracy across all 40 requests when using GPT-5.1, successfully generating correct scripts for all task categories. When using GPT-5-mini, the overall accuracy decreases to 90\%, with most errors occurring in task categories that require more complex parameter extraction beyond single-call requests. For the DeepSeekV3.2 series, DeepSeekV3.2-Reasoner maintains an overall accuracy of 95\%, while DeepSeekV3.2-Chat achieves 77.5\%. These results indicate that more complex request types such as multi-call, testbench auto-design, and PPA optimization place higher demands on parameter parsing fidelity and rule adherence.

For LLM back ends other than GPT-5.1, further improvements are feasible by refining agent prompts with more comprehensive rule coverage or by introducing additional verification and calibration agents to cross-check scheduling decisions and extracted parameters. Recent work has shown that multi-agent frameworks with more structured and fine-grained workflows can significantly improve robustness and reliability in complex EDA tasks \cite{bg_verilogcoder,exp_zhao2025mage}. In this work, we take an initial step toward systematic evaluation of LLM-based agents for CIM workflows and demonstrate that ChatNeuroSim can reliably process all evaluated requests when incorporated with GPT-5.1. Reducing agent execution cost and improving efficiency under alternative LLM back ends are left for future work.

\subsection{Case Study 2: FoM Optimization on ResNet-50 and Swin-T without Design Space Pruning} \label{sec:casestudy2}

\begin{figure}[h]
  \centering
  \includegraphics[width=\linewidth]{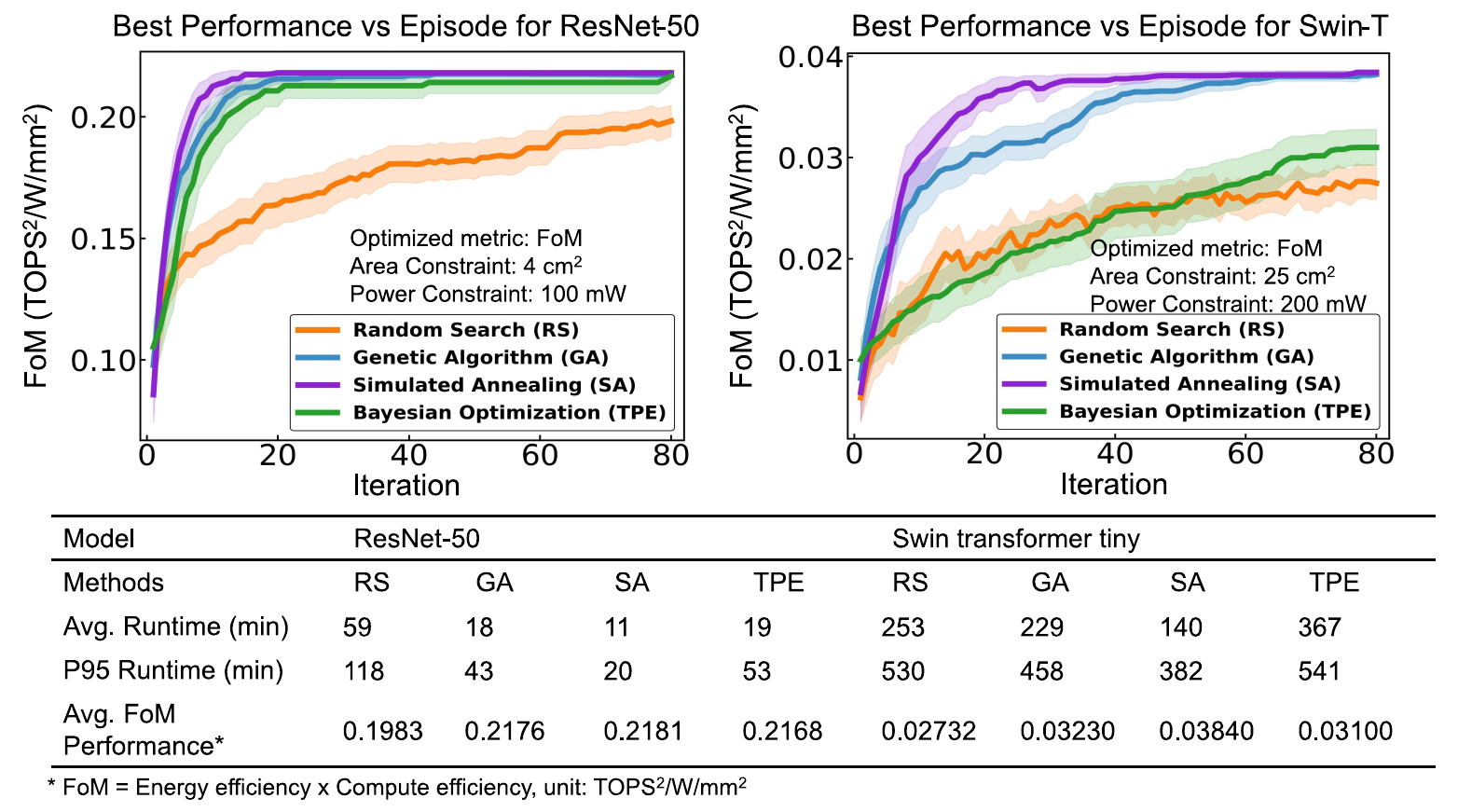}
  \caption{Case study 2: Best FoM performance versus iteration using different optimization strategies for (a) ResNet-50 and (b) Swin-T without design space pruning. The shaded area shows the 95\% bootstrap confidence interval over 50 runs.}
  \label{casestudy2}
\end{figure}

In case study 2, we evaluate the optimized performance and runtime of the proposed CIM optimizer without design space pruning. The goal is to establish a baseline and answer the following question:
\emph{Are conventional heuristic optimization algorithms sufficient to identify optimal PPA configurations with fast convergence for different DNN workloads?}

We conduct Figure-of-Merit (FoM) optimization on both ResNet-50 and Swin Transformer Tiny (Swin-T) under power and area constraints using four optimization algorithms: Random Search (RS), Genetic Algorithm (GA), Simulated Annealing (SA), and Bayesian optimization with Tree-Structured Parzen Estimator (TPE). FoM is defined as the product of energy efficiency and compute efficiency. Each experiment runs for 80 optimization iterations and is repeated over 50 independent runs.

Fig. \ref{casestudy2} reports the average best FoM achieved by each algorithm, along with the corresponding runtime statistics. For ResNet-50, all three heuristic algorithms (GA, SA, and TPE) converge faster than Random Search. Among them, SA consistently exhibits the fastest convergence, achieving superior average FoM performance with the lowest average runtime and P95 runtime. Here, runtime is defined as the wall-clock time of the iteration at which the optimizer first identifies the best-performing configuration. All runtime values are estimated using the characterization-based cost model described in Section \ref{sec:setup}. The P95 runtime is computed from the empirical distribution over 50 runs and represents the runtime within which 95\% of runs complete, providing an estimation of near-worst-case performance.

As summarized in Fig. \ref{casestudy2}, ResNet-50 optimization completes within approximately one hour using GA, SA, or TPE. In contrast, Swin-T optimization exhibits significantly longer runtime due to both increased NeuroSim simulation latency and a larger design space. Even with SA, the most efficient algorithm among the evaluated methods, the optimizer requires several hours on average, and up to 6–7 hours in the worst 5\% of runs to identify the optimal FoM configuration. These results indicate that while existing heuristic algorithms are effective for CNN-based workloads, their runtime overhead becomes prohibitive for vision transformer optimization. This observation motivates the introduction of design space pruning techniques tailored for transformer-based CIM optimization, which are evaluated in the subsequent case studies.

\begin{itemize}
\item \emph{Key takeaway:} Vision transformer workloads (e.g., Swin-T) incur higher optimization runtime than convolutional neural networks (e.g., ResNet-50) due to larger design spaces and longer simulation latency. Algorithm-level enhancements beyond standard heuristic optimization are  necessary to improve convergence speed for transformer PPA optimization.
\end{itemize}

\subsection{Case Study 3: Optimizations on Swin-T with and without Design Space Pruning under Different Algorithms}
\begin{figure}[h]
  \centering
  \includegraphics[width=\linewidth]{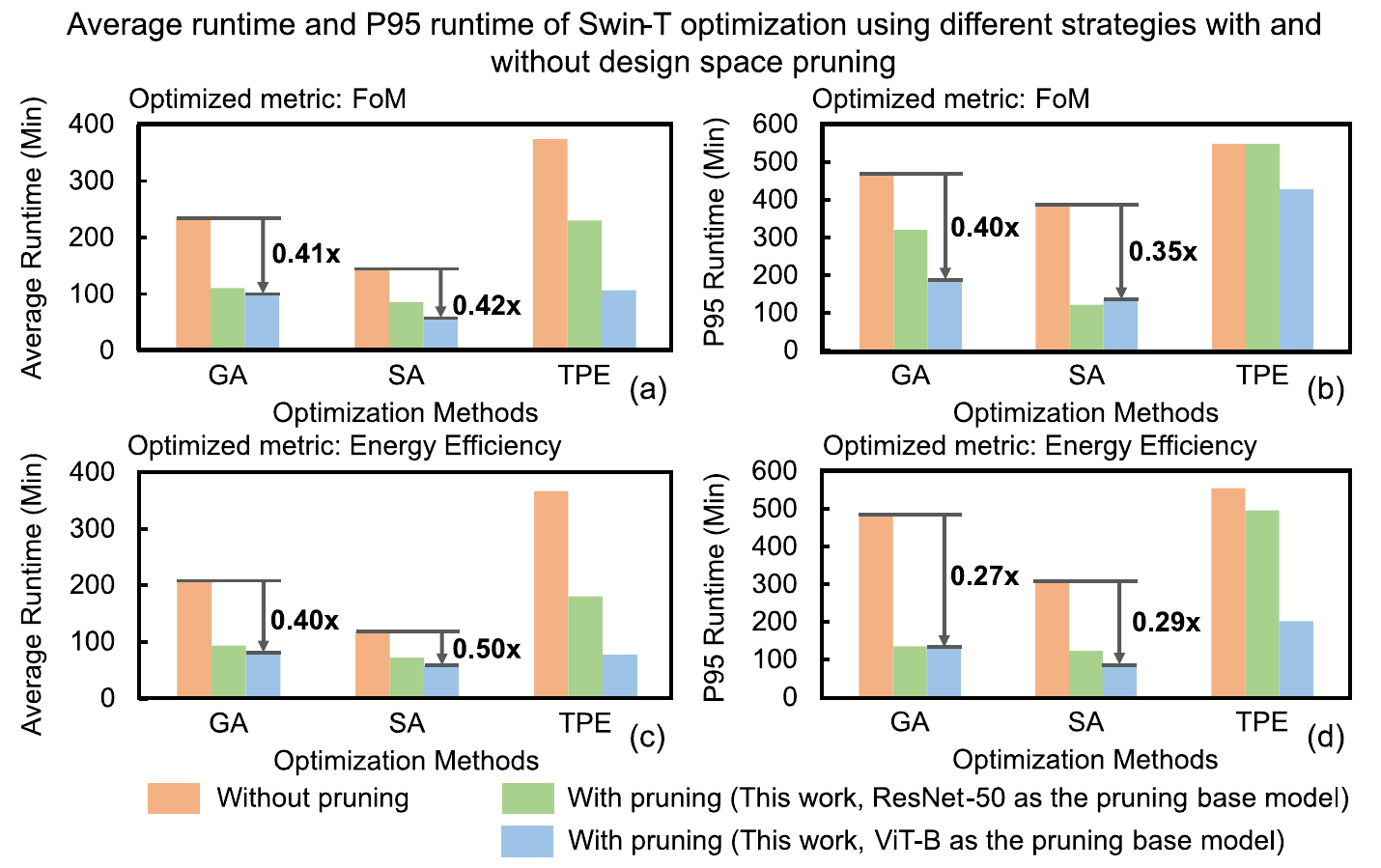}
  \caption{Case study 3: Average runtime (a,c) and P95 runtime (b,d) of Swin-T optimization using different strategies with and without design space pruning. (Area constraint: 25 cm$^2$, Power constraint: 200 mW, Optimization iterations: 80)}
  \label{casestudy3_1}
\end{figure}

In this case study, we evaluate the effectiveness of the proposed design space pruning strategy for Swin Transformer Tiny (Swin-T) optimization. We compare the optimization runtime and performance of FoM and energy efficiency using three heuristic algorithms (i.e., GA, SA, and TPE) with and without design space pruning. When pruning is applied, two base models are considered: ResNet-50 and Vision Transformer Base (ViT-B). For each base model, we assume the PPA of the corresponding design space is pre-simulated and stored in a database. During initial pruning (Algorithm \ref{alg:topk-pruning}), the optimizer analyzes the performance distribution of the base model and prunes the Swin-T design space accordingly before the optimization starts.

Fig. \ref{casestudy3_1} reports the optimization runtime of FoM and energy efficiency. For FoM optimization, design space pruning based on ViT-B reduces the average runtime by 0.42x and the P95 runtime by 0.35x using simulated annealing, compared to the baseline without pruning. Similarly, for energy efficiency optimization, the same pruning strategy achieves a 0.50x reduction in average runtime and a 0.29x reduction in P95 runtime. Across all heuristic algorithms, SA consistently shows the fastest convergence speed, both with and without design space pruning, outperforming GA and TPE in terms of average runtime and $P95$ runtime. Comparing the two base models, design space pruning based on ViT-B exhibits consistent runtime speedups across all three heuristic algorithms for both optimization objectives. In contrast, the pruning based on ResNet-50 yields limited runtime improvement for TPE over the no-pruning baseline. The reason behind this behavior could be (1) differences in candidate sampling strategies across algorithms and (2) local stagnation in high-performing regions after early convergence within the pruned design space.

\begin{figure}[h]
  \centering
  \includegraphics[width=\linewidth]{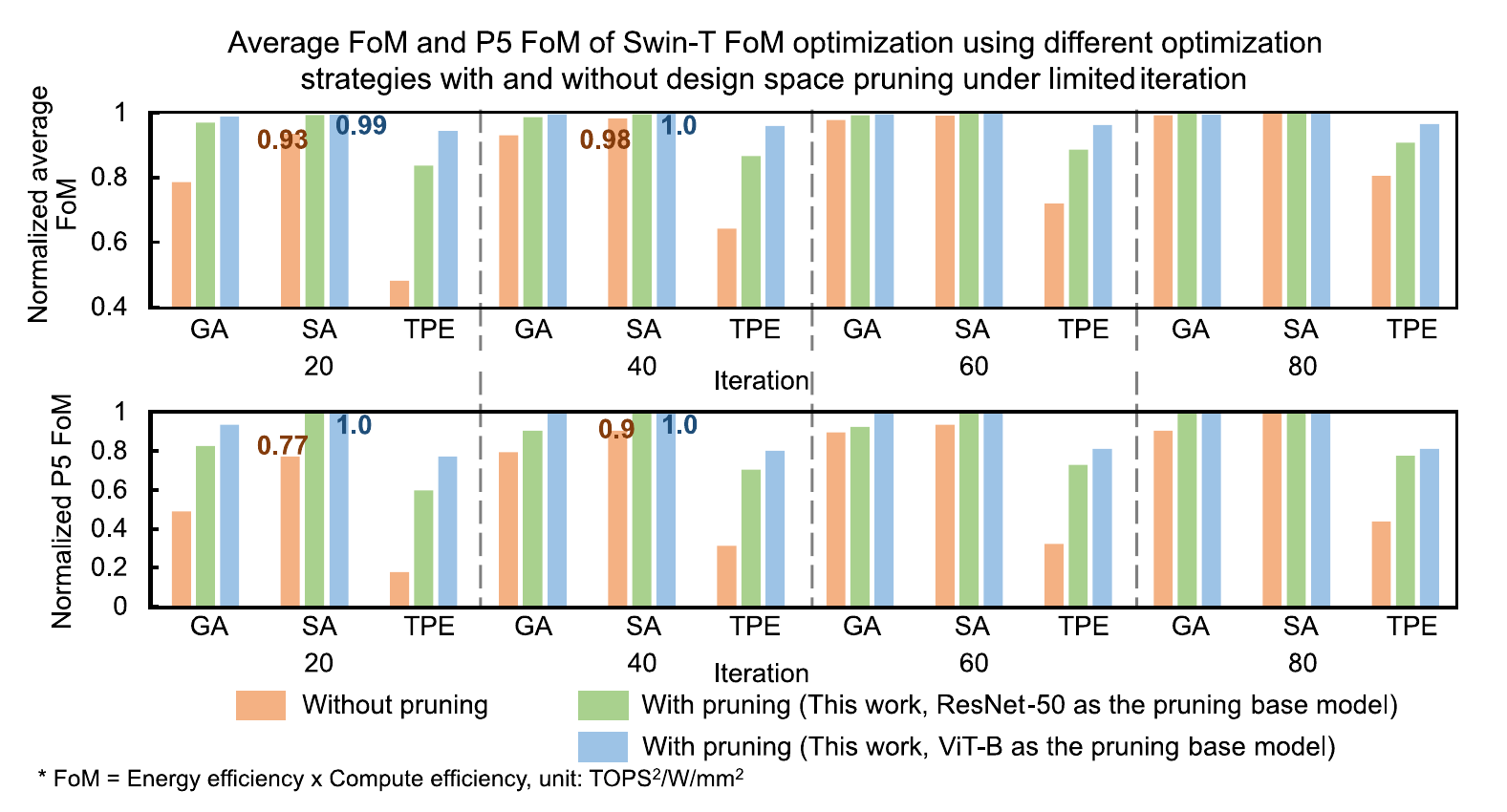}
  \caption{Case study 3: Average FoM and P5 FoM of Swin-T FoM optimization using different optimization strategies with and without design space pruning under limited iteration. (Area constraint: 25 cm$^2$, Power constraint: 200 mW)}
  \label{casestudy3_2}
\end{figure}

We further evaluate the quality of the optimized performance metric within a fixed iteration budget. The FoM values are normalized to the global optimum. We report both the average FoM and the P5 FoM, where P5 FoM indicates that 95\% of optimization runs achieve a FoM value higher than this threshold. As shown in Fig. \ref{casestudy3_2}, design space pruning based on either ResNet-50 or ViT-B consistently yields higher average FoM and P5 FoM compared to the baseline without pruning across all algorithm–iteration configurations. Among the three algorithms, SA achieves the highest optimized performance under the 20 and 40 iteration limit. Notably, although TPE with design space pruning does not always demonstrate significant runtime reduction (Fig. \ref{casestudy3_1}), it consistently identifies higher-quality solutions than the corresponding no-pruning baseline. These results indicate that design space pruning not only accelerates optimization in these cases but also improves solution quality when optimization iterations are limited. The key takeaways from this case study are:

\begin{itemize}
\item \emph{Simulated annealing consistently achieves lower average and $P95$ runtime than GA and TPE for both FoM and energy efficiency optimization, regardless of whether pruning is applied.}
\item \emph{Design space pruning based on ViT-B provides consistent runtime speedups across all evaluated algorithms and objectives compared to optimization without pruning.}
\item \emph{Design space pruning based on both ResNet-50 and ViT-B improves average and P5 FoM under limited iteration budgets for all three heuristic algorithms.}
\end{itemize}

\subsection{Case Study 4: FoM Optimizations on Swin-T with and without Design Space Pruning under Different Hardware Constraints}

\begin{figure}[h]
  \centering
  \includegraphics[width=\linewidth]{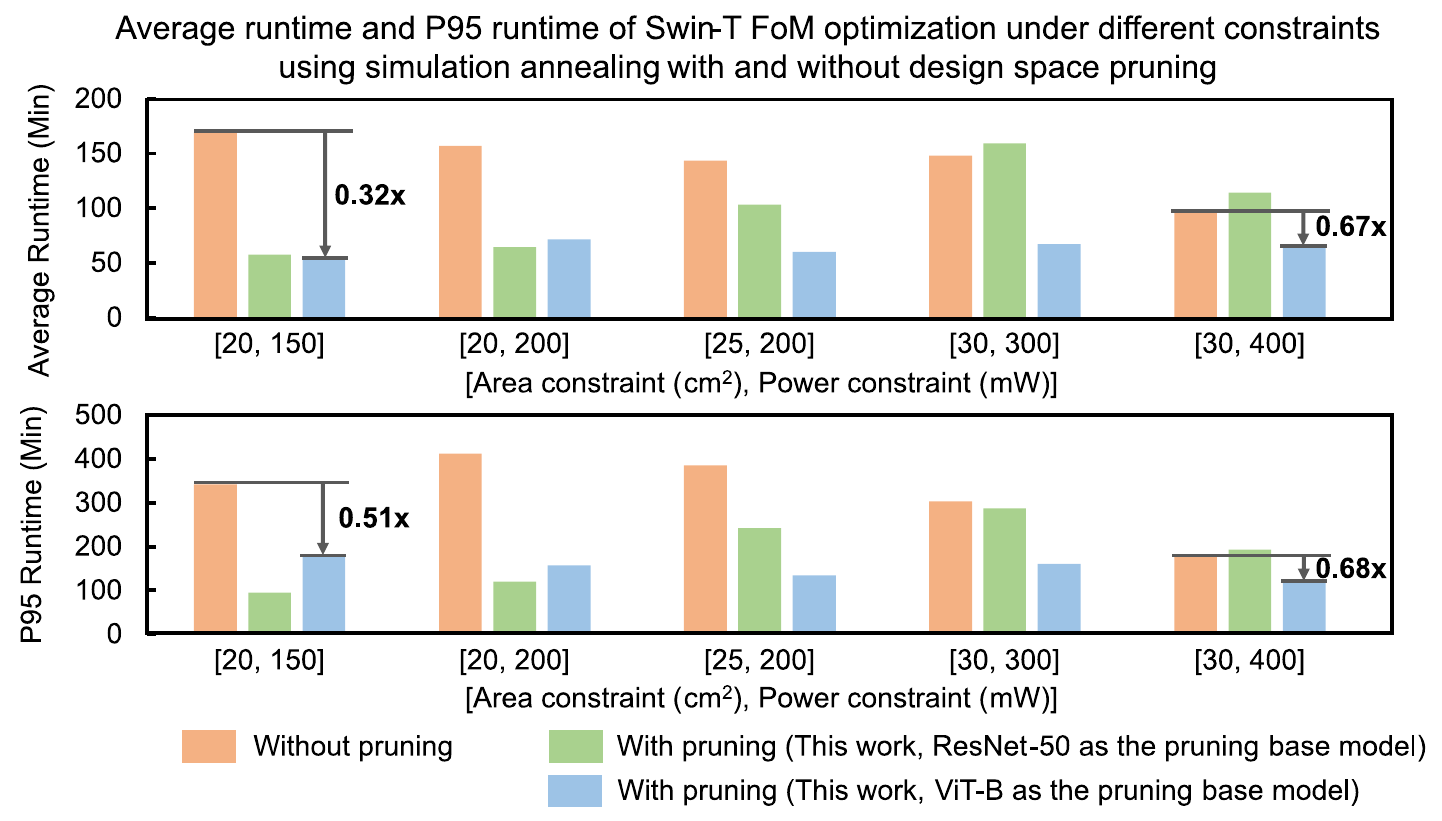}
  \caption{Case study 4: Average runtime and P95 runtime of Swin-T FoM optimization under different constraints using simulation annealing with and without design space pruning. (Optimization iterations: 80)}
  \label{casestudy4}
\end{figure}

As simulated annealing (SA) demonstrates higher search efficiency compared to genetic algorithm (GA) and Tree-Structured Parzen Estimator (TPE), we adopt SA as the optimization algorithm in the following three case studies. Case Study 4 evaluates the generality of the proposed design space pruning strategy when hardware constraints are scaled. Fig. \ref{casestudy4} reports the average runtime and P95 runtime of Swin-T FoM optimization under five different power and area constraint configurations. When the hardware constraints are relaxed, the baseline optimization without pruning converges faster, as a larger fraction of the design space becomes feasible and high-performing configurations are easier to identify.

For design space pruning based on ResNet-50, runtime improvement is observed under tighter constraints. However, under looser constraints, pruning may yield limited benefits or even degrade runtime performance. This behavior can be attributed to several factors: (1) reduced accuracy of constraint projection when extrapolating to significantly relaxed constraints, (2) early stagnation within locally high-performing regions of the pruned design space, and (3) additional verification overhead introduced by the stochastic de-pruning mechanism (Algorithm \ref{alg:depruning}) during optimization.

In contrast, design space pruning based on ViT-B consistently achieves runtime reduction across all evaluated constraint configurations. This observation indicates a high degree of similarity between the FoM-relevant design spaces of Swin-T and ViT-B. Compared to the no-pruning baseline, design space pruning based on ViT-B achieves an average runtime reduction of 0.32x–0.67x and a P95 runtime reduction of 0.51x–0.68x across the five constraint configurations.

\begin{itemize}
\item \emph{Key takeaway:} Design space pruning based on ViT-B consistently accelerates Swin-T FoM optimization across a wide range of hardware constraints, demonstrating the generality and robustness compared to optimization without pruning.
\end{itemize}

\subsection{Case Study 5: Different Metric Optimizations on Swin-T with and without Design Space Pruning}
\begin{figure}[h]
  \centering
  \includegraphics[width=\linewidth]{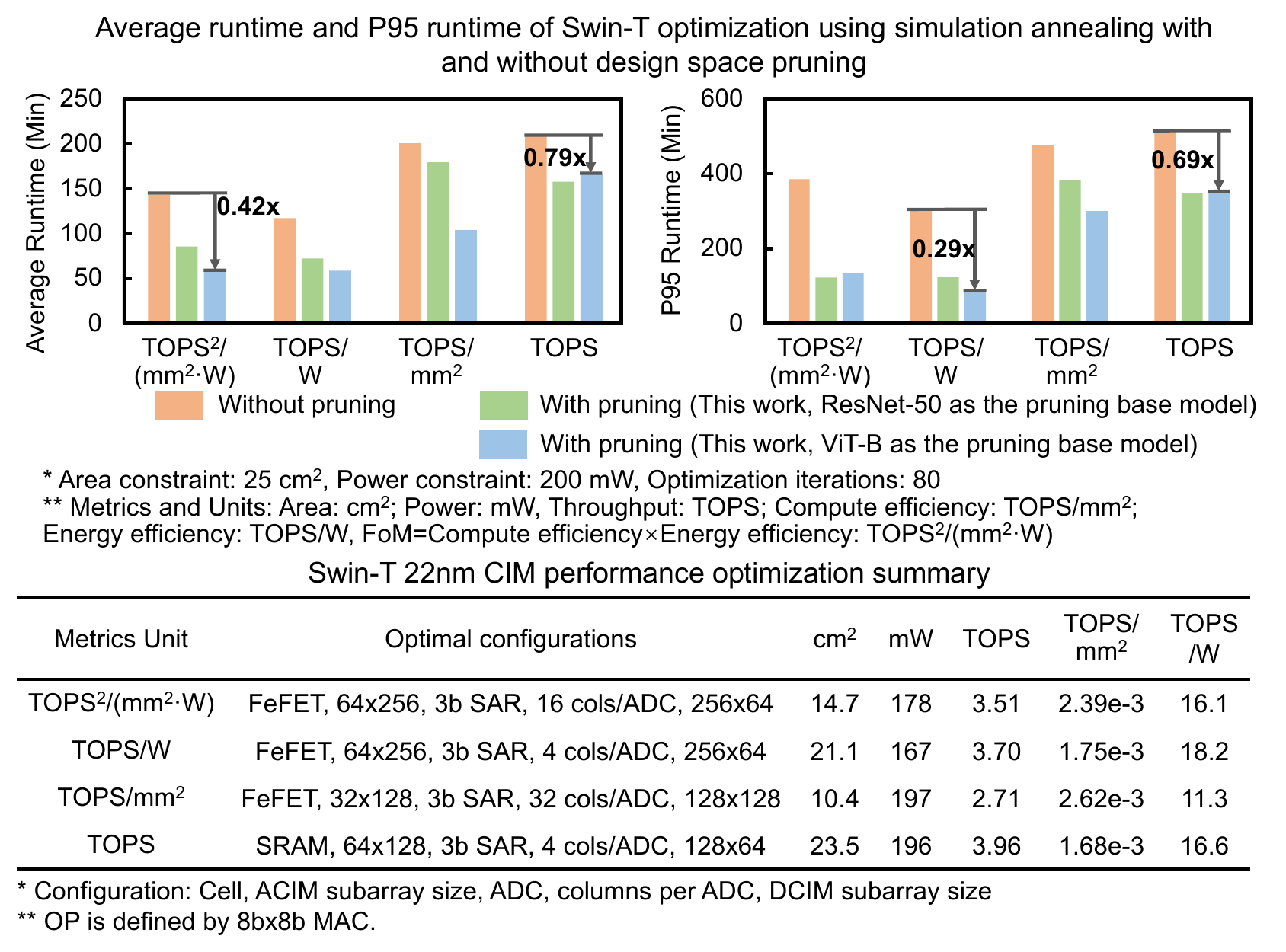}
  \caption{Case study 5: Average runtime and P95 runtime of Swin-T optimization using simulation annealing with and without design space pruning, and Swin-T 22nm CIM performance optimization summary.}
  \label{casestudy5}
\end{figure}

In this case study, we evaluate the generality of the proposed design space pruning strategy across different optimization objectives. Specifically, we consider four commonly used performance metrics for CIM systems: FoM, energy efficiency, compute efficiency, and throughput. All experiments are conducted on Swin Transformer Tiny (Swin-T) using simulated annealing with and without design space pruning.

Fig. \ref{casestudy5} reports the optimization runtime of each metric. For all four objectives, integrating design space pruning based on either ResNet-50 and Swin-T consistently reduces the runtime required to identify the optimal configurations. In particular, design space pruning based on ViT-B achieves a 0.42x–0.79x reduction in average runtime and a 0.29x–0.69x reduction in P95 runtime compared to the no-pruning baseline. The summary in Fig. \ref{casestudy5} further lists the optimal 22 nm CIM configurations and the corresponding performance metrics obtained for Swin-T under each optimization objective. These results demonstrate that the proposed pruning framework generalizes effectively beyond single-metric optimization, enabling efficient design space exploration across diverse CIM performance criteria.

\begin{itemize}
\item \emph{Key takeaway:} Design space pruning based on either ResNet-50 and ViT-B consistently accelerates Swin-T optimization across multiple performance objectives, demonstrating metric-level generality of the proposed optimization framework.
\end{itemize}

\subsection{Case Study 6: Hyper-parameter Design Space of Design Space Pruning}
\begin{figure}[h]
  \centering
  \includegraphics[width=\linewidth]{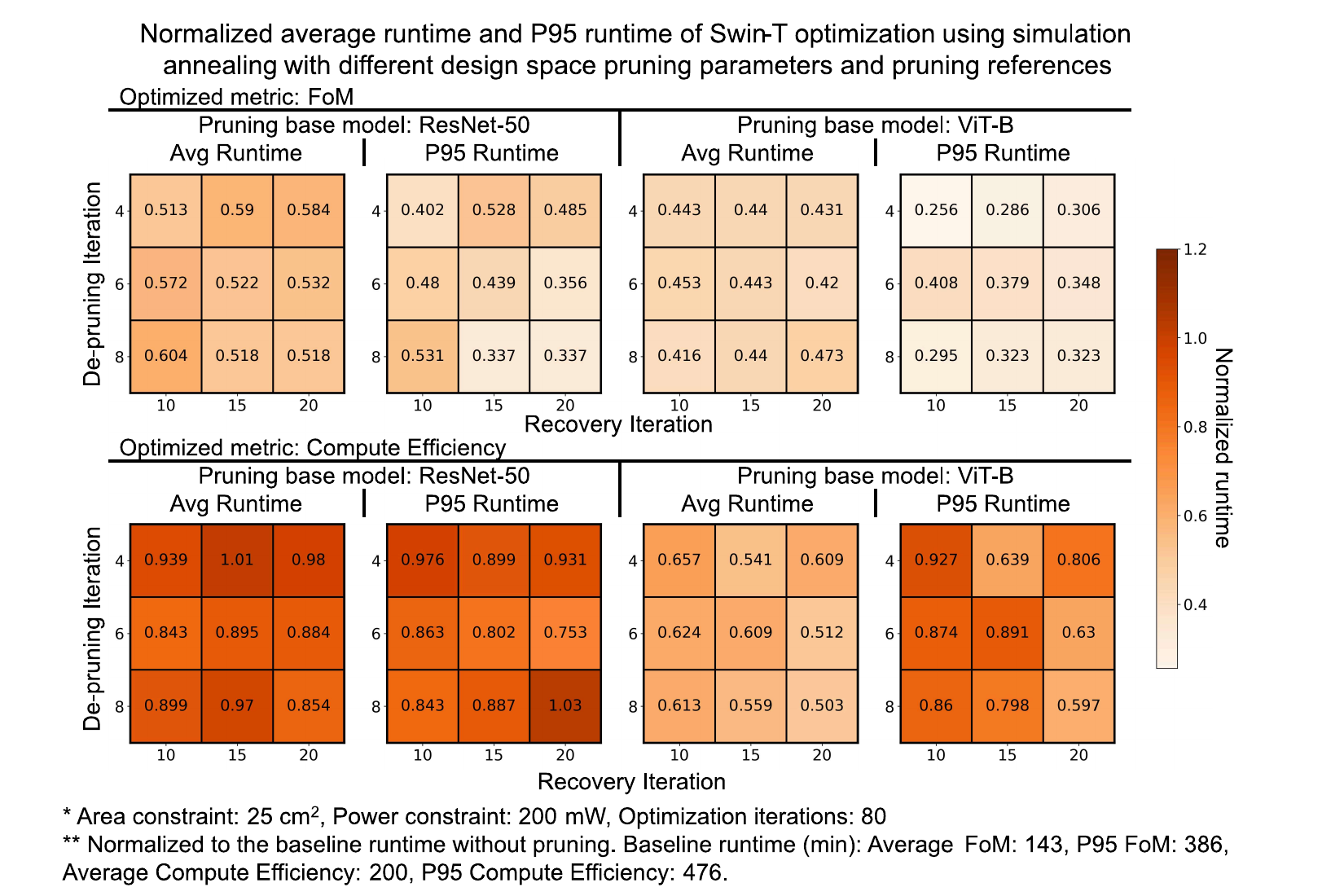}
  \caption{Case study 6: Normalized average runtime and P95 runtime of Swin-T optimization using simulation annealing with different design space pruning parameters and pruning base models.}
  \label{casestudy6}
\end{figure}

In the final case study, we analyze the impact of hyper-parameter choices in the proposed design space pruning framework on optimization runtime. Fig. \ref{casestudy6} reports the average runtime and P95 runtime for Swin-T FoM and compute-efficiency optimization, normalized to the corresponding baseline runtime without pruning. We focus on two key hyper-parameters: the \emph{de-pruning iteration} and the \emph{recovery iteration} (Fig. \ref{optimizerFramework}). The de-pruning iteration specifies the iteration at which stochastic de-pruning is terminated. Prior to this point, the de-pruning algorithm is invoked at fixed intervals (e.g., every two iterations in Case Studies 2–6) to gradually restore excluded parameter values. After de-pruning stops, the optimizer continues searching within the most recently de-pruned design space until the recovery iteration is reached, at which point the full original design space is restored.

For FoM optimization with design space pruning based on ResNet-50, increasing the de-pruning iteration from 4 to 8 under a recovery iteration of 10 leads to higher runtime. This behavior indicates that the optimizer fails to identify the optimal configuration within the pruned space during the de-pruning stage and quickly transitions to searching the full design space with insufficient pruned space search. Moreover, extending the de-pruning phase incurs additional verification overhead due to extra simulations introduced by the de-pruning process.

Similarly, increasing the recovery iteration from 10 to 15 under a de-pruning iteration of 4 results in longer runtime, as the optimizer continues to explore a pruned space that excludes the optimal configuration for an extended period. In contrast, under larger de-pruning iterations (6 or 8), increasing the recovery iteration reduces runtime because the pruned design space has been sufficiently verified and adjusted during de-pruning.

Among all evaluated configurations based on ResNet-50, a combination of 4 de-pruning iterations and 10 recovery iterations yields the largest average runtime reduction compared to the no-pruning baseline. This configuration corresponds to an aggressive pruning strategy, where the optimizer strongly exploits knowledge transferred from the base model during early iterations and rapidly recovers the full design space if the transferred experience proves ineffective. Conversely, a configuration with 8 de-pruning iterations and 20 recovery iterations achieves the lowest P95 runtime. This conservative strategy emphasizes sufficient verification of transferred knowledge before committing to prolonged exploration within the pruned space, thereby reducing near-worst-case runtime.

When optimizing compute efficiency, the benefits of design space pruning are less significant than for FoM optimization. With ResNet-50 as the base model, the maximum average runtime reduction is 0.843x and the maximum P95 runtime reduction is 0.802x. In some configurations, runtime even exceeds the no-pruning baseline. This behavior suggests that, for compute efficiency, the pruned design space derived from ResNet-50 often does not contain the optimal Swin-T configuration due to differences in search-space characteristics. In addition, the verification overhead introduced by de-pruning further compromises potential runtime gains.

In contrast, using ViT-B as the base model yields more improvements, achieving 0.503x average runtime reduction and 0.597x P95 runtime reduction. With recovery iteration set to 20 and de-pruning iteration set to 6 or 8, the runtime can be effectively reduced compared to the no-pruning baseline. These results indicate a higher degree of design-space similarity between ViT-B and Swin-T, providing effective knowledge transfer across different performance objectives.

\subsection{Summary of Case Study 2-6 and Design Guidelines}

Based on the experimental results from Case Study 2-6, we summarize the following practical guidelines for applying design space pruning in CIM optimization:

\begin{itemize}
    \item \emph{For convolutional neural networks (e.g., ResNet-50), optimization runtime is typically short, and design space pruning is not strictly necessary. For vision transformers (e.g., Swin-T), which involve larger design spaces and longer simulation time, design space pruning is highly effective in accelerating optimization.}
    \item \emph{Simulated annealing consistently provides faster convergence than genetic algorithm and Tree-Structured Parzen Estimator, both with and without design space pruning.}
    \item \emph{An optimization budget of approximately 40 iterations is suggested under 32-core parallelism based on empirical evaluation. Note that this value may be adjusted depending on available compute resources and time constraints.}
    \item \emph{For vision transformer optimization, selecting a base model with similar architectural characteristics (e.g., ViT-B for Swin-T) is critical for effective design space pruning across different hardware constraints and optimization objectives.}
    \item \emph{The de-pruning and recovery iterations significantly influence optimization runtime. These parameters require careful design based on the desired trade-off between exploration efficiency and robustness.}
\end{itemize}

All of these guidelines are integrated into ChatNeuroSim to assist users in selecting appropriate optimization strategies when targeting CIM PPA under hardware constraints. By automating simulator interaction, parameter management, and optimization with adaptive design space pruning, ChatNeuroSim substantially reduces CIM design effort and provides rapid identification of high-performing hardware configurations.

\section{Conclusion}

This work presents ChatNeuroSim, an LLM-based agent framework for automated CIM accelerator deployment and optimization. The proposed framework supports four types of CIM requests through a coordinated workflow of task parsing, parameter parsing, and parameter adjustment agents. The correctness and robustness of the agent system are validated using a customized CIM request dataset, achieving a 100\% pass rate in one-shot script generation. In addition, we introduce a CIM optimizer with adaptive design space pruning that transfers and verifies optimization experience from prior base models when optimizing the PPA of new target workloads. Across extensive case studies, the proposed design space pruning strategy achieves a 0.42x–0.79x reduction in average optimization runtime and a 0.29x–0.69x reduction in P95 runtime compared to baseline optimization without design space pruning. We further analyze the effects of optimization objectives, hardware constraints, base-model selection, and pruning hyper-parameters, and integrate the resulting design guidelines into ChatNeuroSim.

By automating simulator interaction, parameter dependency handling, and PPA optimization, ChatNeuroSim reduces the manual effort required to interpret simulator documentation and manage complex design configurations. The proposed design space pruning technique further accelerates design space exploration by improving search efficiency without sacrificing solution quality. The ChatNeuroSim framework and optimization methodology significantly reduce the manpower and time in CIM design space exploration, facilitating rapid and efficient development of high-performance CIM accelerators. ChatNeuroSim UI is open-sourced to the research community with trial runs available to interested users (\url{https://github.com/neurosim/ChatNeuroSim}).


\bibliographystyle{ACM-Reference-Format}
\bibliography{samples/references}

\newpage
\appendix
\section*{Appendix}
\section{Algorithms of Design Space Pruning} \label{appendixDSAlgorithms}
The workflow of three algorithms are demonstrated in Fig. \ref{optimizerFramework}, while the implementation details are shown in this section.

\subsection{Algorithm: Cross-Space Constraint Projection via Power-Law Fitting}
\begin{algorithm}[h]
\textsl{}\setstretch{0.7}
\caption{Cross-Space Constraint Projection via Power-Law Fitting}
\label{alg:constraint-projection}
\SetAlgoLined
\KwIn{
Base design space $\mathcal{D}_b$, target design space $\mathcal{D}_t$; 
Base dataset $\mathcal{S}_b$; 
Constraint set $\mathcal{C}$; 
Sample budget $N$.
}
\KwOut{
Design space intersection $\mathcal{I}$; 
Fitting coefficients $\{(a_{0}^{m,c}, a_{1}^{m,c})\}$ for all memory types $m \in \mathcal{M}$ and constraints $c \in \mathcal{C}$.
}

$\mathcal{I} \leftarrow \textsc{Intersection}(\mathcal{D}_b, \mathcal{D}_t)$ \tcp*{Step 1: Select shared parameter space}
$\mathcal{M} \leftarrow \mathcal{I}[\text{memCellType}]$ \tcp*{Common memory cell types}

\BlankLine
\tcp{Extract boundary values for representative sampling}
$\mathcal{V} \leftarrow \{ \textsc{MinMax}(\mathcal{I}[k]) \mid k \in \{\text{rowACIM, colACIM, typeADC, levelADC, muxColADC, rowDCIM, colDCIM}\} \}$\;
\BlankLine
\ForEach{$m \in \mathcal{M}$}{
    \tcp{Step 2: Generate candidate sample set $\Omega_m$ based on all parameter combinations in $\mathcal{V}$ }
    $\Omega_m \leftarrow \{ (m, r^a, c^a, t, l, u, r^d, c^d) \mid r \in \{V[rowACIM]_{min}, V[rowACIM]_{max} \}, \dots \}$\;
    
    $\mathcal{T}_m, \mathcal{B}_m \leftarrow \textsc{SelectSamples}(\Omega_m, N)$ \tcp*{Random unique subset if $|\Omega_m| > N$}

    \BlankLine
    \tcp{Step 3: Execute performance simulation}
    $\text{Perf}_t \leftarrow \textsc{Simulate}(\mathcal{T}_m)$\;
    $\text{Perf}_b \leftarrow \textsc{Fetch}(\mathcal{S}_b, \mathcal{B}_m)$\;

    \BlankLine
    \ForEach{constraint $c \in \mathcal{C}$}{
        \tcp{Construct pairs $(X_i, Y_i)$ for fitting}
        $\mathbf{X}, \mathbf{Y} \leftarrow \emptyset$\;
        \For{$i = 1$ \KwTo $|\mathcal{T}_m|$}{
            $X_i \leftarrow \text{Perf}_b[\mathcal{B}_m[i]][c]$\;
            $Y_i \leftarrow \text{Perf}_t[\mathcal{T}_m[i]][c]$\;
            $\mathbf{X} \leftarrow \mathbf{X} \cup \{X_i\}, \mathbf{Y} \leftarrow \mathbf{Y} \cup \{Y_i\}$\;
        }
        \BlankLine
        \tcp{Step 4: Linear regression fitting in log-log space}
        $(a_{0}^{m,c}, a_{1}^{m,c}) \leftarrow \textsc{LeastSquares}(\log(\mathbf{X}), \log(\mathbf{Y}))$\;
    }
}
\Return $\mathcal{I}, \{(a_{0}^{m,c}, a_{1}^{m,c})\}$\;
\end{algorithm}

Algorithm \ref{alg:constraint-projection} is designed to address the challenge of mapping performance constraints from a base design space $\mathcal{D}_b$ to a target design space $\mathcal{D}_t$ when architectural disparities exist. By establishing an analytical cross-model mapping, the algorithm projects the constraints (e.g., power and area) of the base design space to the target design space. Therefore, design points in the base model whose projected performance violates the user-defined constraints are filtered out. The remaining design points are then carried forward for further analysis in the next design space pruning stage.

The algorithm first invokes the Intersection function to identify shared parameter dimensions between the two spaces (e.g., memory cell types, array dimensions, ADC resolutions), denoted as the intersection space $\mathcal{I}$. To ensure the robustness of the fitting model across the design range, the algorithm extracts boundary values (minimum and maximum) for each dimension to construct a parameter set $\mathcal{V}$. Subject to a sampling budget $N$ ($N=32$ in Case Study 3-6), a subset $\mathcal{T}_m$ is randomly drawn from these boundary combinations to cover the design points for projection construction.

Next, for the selected sample points, the algorithm executes different PPA acquisition ($X_i$ and $Y_i$) of two models. For the target model, the \emph{Simulate} function is called to run NeuroSim simulations in the target environment. For the base model, we assume the PPA of design space are pre-simulated before the target model optimization. The corresponding performance records $\text{Perf}_b$ are directly retrieved from the existing base dataset $\mathcal{S}_b$.

Lastly, given that hardware performance metrics typically follow non-linear scaling laws across different technologies, a Power-Law Model is employed based on the equation:

\begin{equation}
    Y = \alpha X^{\beta}
\end{equation}
where $\alpha$ and $\beta$ are space-specific scaling factors.

\begin{equation}
    \ln(Y^{m,c}) = a_1^{m,c} \ln(X^{m,c}) + a_0^{m,c}
\end{equation}
$Y^{m,c}, X^{m,c}$ refers to the value of hardware constraint $c$ and memory type $m$ in target model and base model, respectively. The LeastSquares method is applied to the transformed data to perform linear regression. The resulting projection coefficients are used in subsequent design space exploration stages to rapidly project constraints from the base space to guide the search in the target space.

\subsection{Algorithm: Top-K Design Space Pruning}

\begin{algorithm}[h]
\setstretch{0.7}
\caption{Top-K Design Space Pruning}
\label{alg:topk-pruning}
\SetAlgoLined
\KwIn{
    Optimization target $T_{opt}$; 
    Shared intersection design space $\mathcal{I}$;
    Base dataset $\mathcal{S}_b$;
    Constraint set $\mathcal{C}$; 
    Fitting coefficients $\{(a_{0}^{m,c}, a_{1}^{m,c})\}$; 
    Hyper-parameters: Top-K ratio $\rho$, Top-bin boundary $\tau$.
}
\KwOut{
    Pruned target design space $\mathcal{D}_t^*$.
}
\BlankLine
\tcp{Step 1: Constraint projection and validity filtering on base design space}
$\Omega_{valid} \leftarrow \emptyset$\;
$\mathcal{S}_{all} \leftarrow \textsc{CartesianProduct}(\mathcal{I})$ \tcp*{Enumerate all shared design points}
\ForEach{$s \in \mathcal{S}_{all}$}{
    $\text{Perf}_s \leftarrow \textsc{Fetch}(\mathcal{S}_b, s)$\;
    $meets\_constraints \leftarrow \text{True}$\;
    \ForEach{constraint $c \in \mathcal{C}$}{
        $m \leftarrow s[memCellType]$\;
        $P_{base} \leftarrow \text{Perf}_s[c]$ \tcp*{Query performance from base dataset $\mathcal{S}_b$}
        $\hat{P}_{target} \leftarrow \exp(a_{0}^{m,c}) \cdot (P_{base})^{a_{1}^{m,c}}$ \tcp*{Apply power-law projection}
        \If{$\hat{P}_{target} > \text{Threshold}_c$}{
            $meets\_constraints \leftarrow \text{False}$\;
            \textbf{break}\;
        }
    }
    \If{$meets\_constraints$}{
        $\Omega_{valid} \leftarrow \Omega_{valid} \cup \{(s, \text{Perf}_s[T_{opt}])\}$\;
    }
}
\BlankLine
\tcp{Step 2: Performance ranking and frequency aggregation}
$K \leftarrow \lceil \rho \cdot |\Omega_{valid}| \rceil$\;
$\Omega_{topk} \leftarrow \textsc{SortByPerformance}(\Omega_{valid}, T_{opt})[1 : K]$\;
$Freq \leftarrow \textsc{CalculateFrequency}(\Omega_{topk})$ \tcp*{Statistical distribution per parameter $p$}
\BlankLine
\tcp{Step 3: Bin-based mapping and pruning}
$\mathcal{D}_t^* \leftarrow \mathcal{D}_t$\;
\ForEach{parameter $p \in \mathcal{I}$}{
    $\mathcal{B}_{base} \leftarrow \textsc{DiscretizeBins}(\mathcal{I}[p])$ \tcp*{Bins: \{small, mid, large\}}
    $BinFreq \leftarrow \textsc{Aggregate}(Freq[p], \mathcal{B}_{base})$\;
    
    $\mathcal{L}_{chosen} \leftarrow \textsc{TopBinSelection}(BinFreq, \tau)$ \tcp*{Select preferred bins}
    
    $\mathcal{B}_{target} \leftarrow \textsc{DiscretizeBins}(\mathcal{D}_t[p], \text{num}=|\mathcal{B}_{base}|)$\;
    $\mathcal{D}_t^*[p] \leftarrow \{ v \mid v \in \mathcal{B}_{target}[l], \forall l \in \mathcal{L}_{chosen} \}$\;
}
\Return $\mathcal{D}_t^*$\;
\end{algorithm}

Algorithm \ref{alg:topk-pruning} leverages the analytical projection coefficients derived in Algorithm \ref{alg:constraint-projection} to prune the large target design space $\mathcal{D}_t$. By projecting the performance of the design points in the base model, the high-performing regions in the base design space are identified within the valid searching space. These pruned regions with reduced search volume are configured as the initial design space during target model optimization.

The algorithm begins by iterating through the Cartesian product of the intersection design space $\mathcal{I}$ to evaluate all shared design points. Under the assumption in Algorithm \ref{alg:constraint-projection}, the base performance $P_{base}$ of each point $s$ is retrieved from the dataset $\mathcal{S}_b$. Using the power-law fitting coefficients $(a_0, a_1)$, the algorithm computes the projected performance $\hat{P}_{target}$ in the target space via the formula:
\begin{equation}
    \hat{P}_{target} = \exp(a_{0}^{m,c}) \cdot (P_{base})^{a_{1}^{m,c}}
\end{equation}
Any design point that violates the projected constraints ($\hat{P}_{target} > \text{Threshold}_c$) is removed. Therefore, the resulting valid set $\Omega_{valid}$ contains only design points likely to satisfy the hardware requirements in the target environment.

Next, from the valid set $\Omega_{valid}$, the algorithm identifies the Top-K performing designs based on the optimization target $T_{opt}$. The hyper-parameter $\rho$ ($\rho=0.2$ in Case Study 3-6) determines the selection ratio. The algorithm then performs a statistical frequency analysis (\emph{CalculateFrequency}) on these elite samples to determine the distribution of each hardware parameter $p$. This process identifies "high-probability" parameter ranges with high performance.

In the last step, to bridge the potential difference in absolute parameter values between the base and target design spaces, the algorithm employs a Bin-based Mapping strategy. The range of each parameter in both spaces is divided into discrete bins (e.g., "small," "medium," "large"). Bins in the base space that exceed a frequency threshold $\tau$ ($\tau=0.85$ in Case Study 3-6) are marked as preferred regions ($\mathcal{L}_{chosen}$). The target design space $\mathcal{D}_t$ is then pruned such that only values within the corresponding preferred bins are retained. The final output $\mathcal{D}_t^*$, is a narrowed design space that focuses on exploring "potential" high-performing regions based on the experience analyzed from the base model.

\subsection{Stochastic Design Space De-pruning via Performance Verification}

\begin{algorithm}[h]
\setstretch{0.7}
\caption{Stochastic Design Space De-pruning via Performance Verification}
\label{alg:depruning}
\SetAlgoLined
\KwIn{
    Current pruned design space $\mathcal{D}_t^*$; 
    Full target design space $\mathcal{D}_t$;
    Constraint set $\mathcal{C}$; 
    Baseline top-performance samples $\mathcal{S}_{base}$;
    Verification budget $N_{total}$;
    Win-rate scaling factor $\gamma$.
}
\KwOut{
    De-pruned design space $\mathcal{D}_{dp}$.
}
\BlankLine
\tcp{Step 1: Identify missing parameter (p) values (v)}
$\mathcal{V}_{miss} \leftarrow \{ (p, v) \mid v \in \mathcal{D}_t[p] \setminus \mathcal{D}_t^*[p], \forall p \in \mathcal{D}_t^* \}$\;
$N_{p} \leftarrow \lfloor N_{total} / |\mathcal{V}_{miss}| \rfloor$ \tcp*{Verification samples per missing value}
\BlankLine
\tcp{Step 2: Monte Carlo verification and performance comparison}
$WinRate \leftarrow \emptyset, \mathcal{P}_{valid} \leftarrow \emptyset$\;
\ForEach{$(p, v) \in \mathcal{V}_{miss}$}{
    $\mathcal{S}_{verify} \leftarrow \emptyset, n_{win} \leftarrow 0$\;
    $\Omega \leftarrow \textsc{SelectSamples}(\mathcal{S}_{base}, N_{p})$\;
    
    \ForEach{$s_b \in \Omega$}{
        $s_{new} \leftarrow \text{ReplaceParam}(s_b, p, v)$\;
        
        $P_{target} \leftarrow \textsc{Simulate}(s_{new})$\;
        \If{$P_{target}$ \text{meets all constraints} $\mathcal{C}$}{
            $\mathcal{P}_{valid} \leftarrow \mathcal{P}_{valid} \cup \{(s_{new}, P_{target})\}$\;
            $P_{base} \leftarrow \text{Performance of } s_b$\;
            \If{$\text{IsBetter}(P_{target}[T_{opt}], P_{base}[T_{opt}])$}{
                $n_{win} \leftarrow n_{win} + 1$\;
            }
        }
    }
    $r_{win} \leftarrow n_{win} / |\Omega|$\;
    $WinRate[p][v] \leftarrow \min(r_{win} \cdot \gamma, 1.0)$ \tcp*{Scale win rate}
}

\BlankLine
\tcp{Step 3: Probabilistic Design Space Restoration}
$\mathcal{D}_{dp} \leftarrow \mathcal{D}_t^*$\;
\ForEach{$(p, v) \in \mathcal{V}_{miss}$}{
    $\alpha \leftarrow \text{UniformRandom}(0, 1)$\;
    \If{$\alpha < WinRate[p][v]$}{
        $\mathcal{D}_{dp}[p] \leftarrow \mathcal{D}_{dp}[p] \cup \{v\}$ \tcp*{Restore missing value}
    }
}

\Return $\mathcal{D}_{dp}$\;
\end{algorithm}

Algorithm \ref{alg:depruning} functions as a corrective mechanism to the pruning performed in Algorithm \ref{alg:topk-pruning}. While the previous pruning reduces the search volume, it could discard high-performance configurations due to the inherent approximations in power-law fitting or bin-based discretization. This algorithm employs a stochastic approach to selectively restore promising parameter values into the design space $\mathcal{D}_t^*$ based on performance verification.

The algorithm first identifies the set of missing parameter values $\mathcal{V}_{miss}$ by calculating the difference between the full target design space $\mathcal{D}_t$ and the current pruned space $\mathcal{D}_t^*$. To manage computational overhead, a total verification budget $N_p$ ($N_{p} = \lfloor N_{total} / |\mathcal{V}_{miss}| \rfloor$, $N_{total}=32$ in Case Study 3-6) is allocated to missing values, defining the number of  trials used to evaluate the potential for restoration.

Secondly, for each candidate value $v$ of parameter $p$, the algorithm conducts a performance comparison against the current top-performing baseline samples $\mathcal{S}_{base}$. A new design configuration $s_{new}$ is generated by substituting the parameter $p$ in a baseline sample $s_b$ with the missing value $v$. After the PPA simulation of $s_{new}$, if $s_{new}$ satisfies all hardware constraints and achieves better performance of $T_{opt}$ compared to the baseline $s_b$, it is recorded as a "win". The final win rate $r_{win}$ represents the empirical probability that the missing value contributes to an optimal design. As the de-pruning process performs periodically, this rate is scaled by the factor $\gamma$ ($\gamma=1.0$ in the initial function call, updated by a constant multiplier 3.0 in subsequent calls) to control the aggressiveness of the restoration.

Finally, the restoration process is determined by a probabilistic threshold. For each missing value, the algorithm generates a uniform random variable $\alpha$. If $\alpha$ is less than the calculated win rate $WinRate[p][v]$, the value is restored to the de-pruned design space $\mathcal{D}_{dp}$. As shown in the flowchart of Fig. \ref{optimizerFramework}, this process can be triggered at specific iteration intervals (de-pruning interval, set to 2 in Case Study 3-6) or when certain recovery thresholds (recovery iteration) are reached. This algorithm adaptively tunes the balance between design space restoration and pruning, dynamically adjusting the searching space with the exposure of high-performing regions.

\section{ChatNeuroSim User Interface (UI) and Dialogue Example} \label{appendixExample}
\begin{figure}[h]
  \centering
  \includegraphics[width=\linewidth]{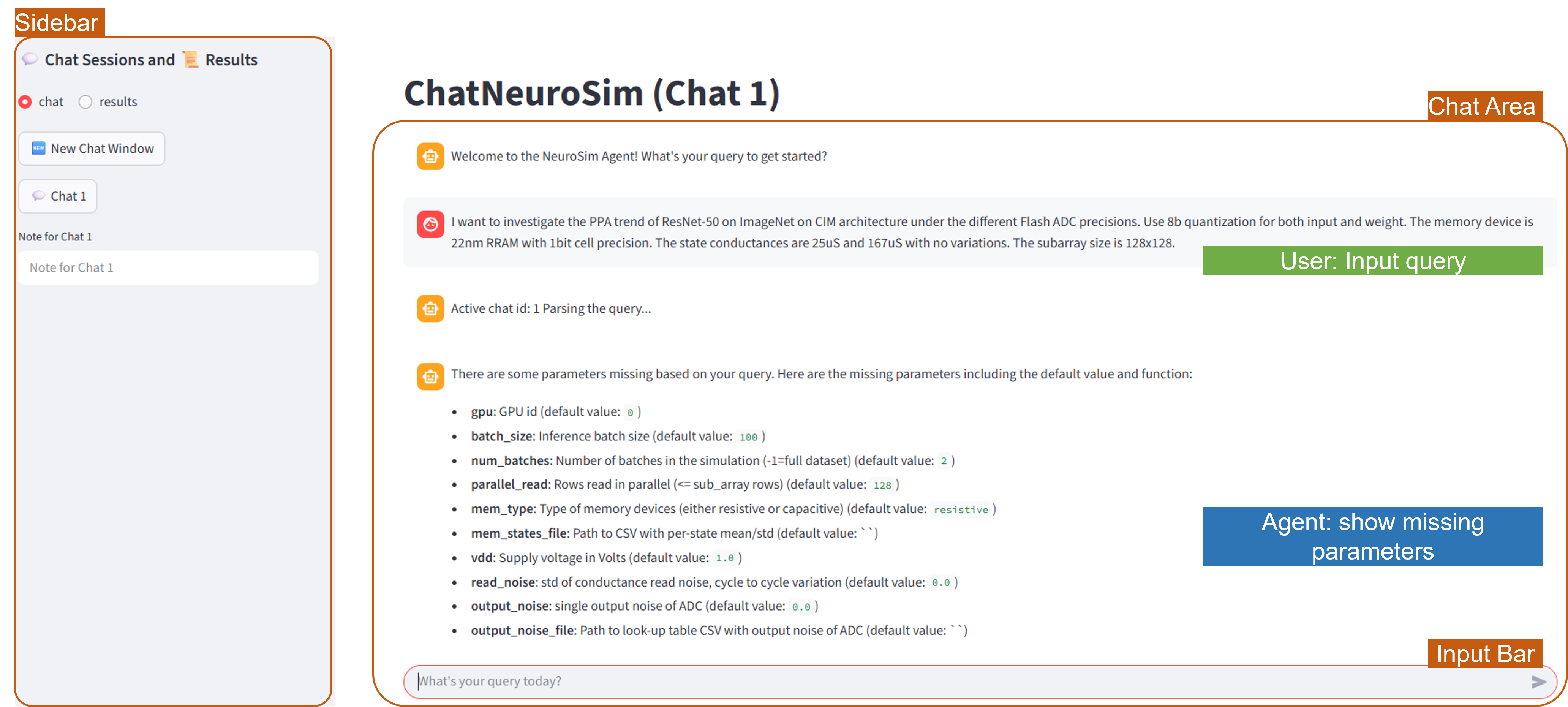}
  \caption{UI overview and dialogue example.}
  \label{appendixDialogue1}
\end{figure}

\begin{figure}[h]
  \centering
  \includegraphics[width=\linewidth]{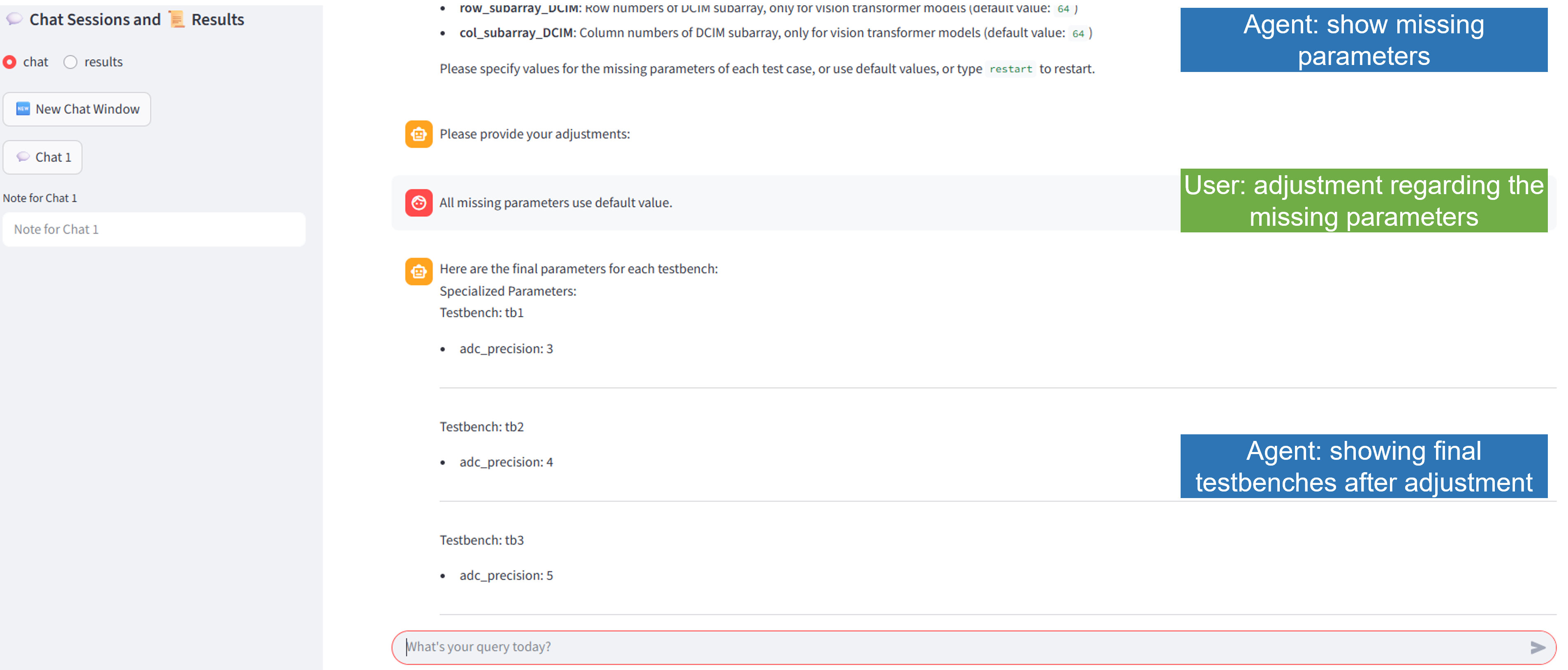}
  \caption{Dialogue example: user adjustment regarding the missing parameters.}
  \label{appendixDialogue2}
\end{figure}

\begin{figure}[h]
  \centering
  \includegraphics[width=\linewidth]{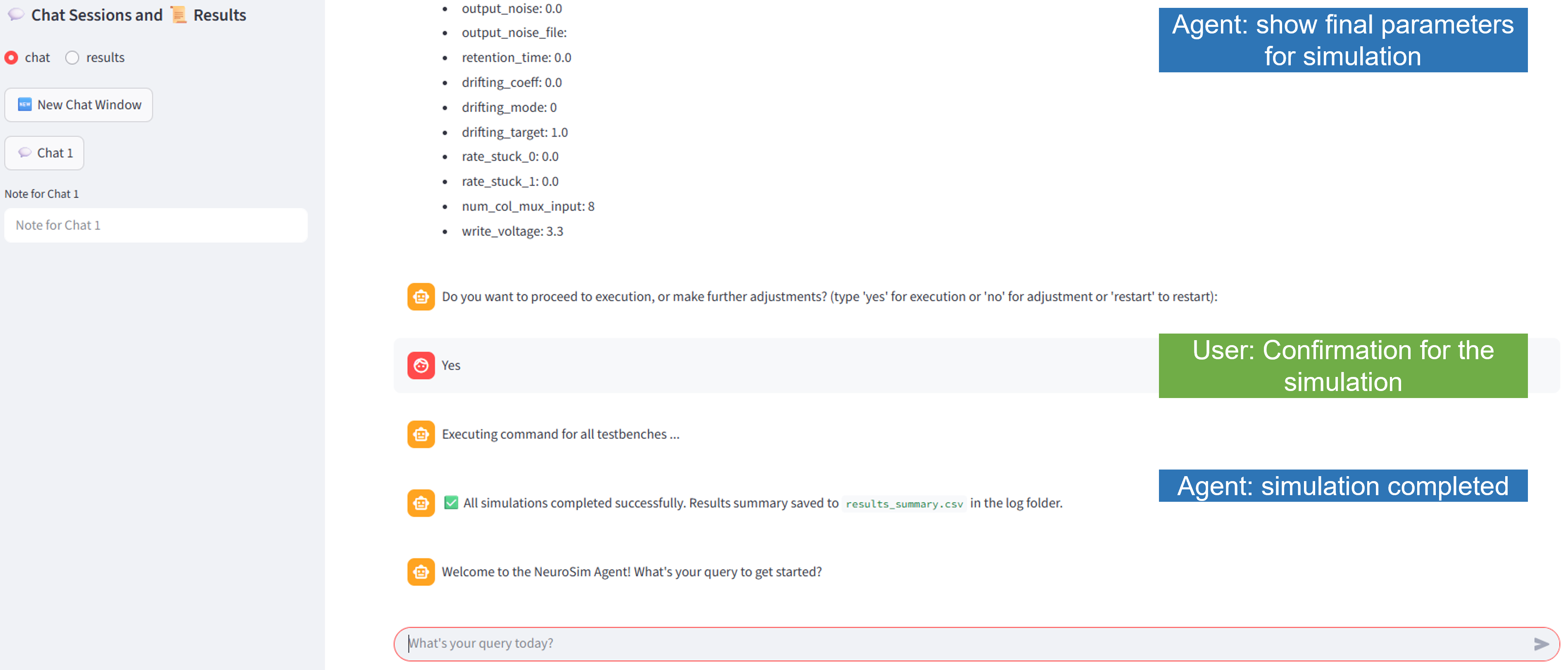}
  \caption{Dialogue example: user confirmation before simulation.}
  \label{appendixDialogue3}
\end{figure}

\begin{figure}[h]
  \centering
  \includegraphics[width=\linewidth]{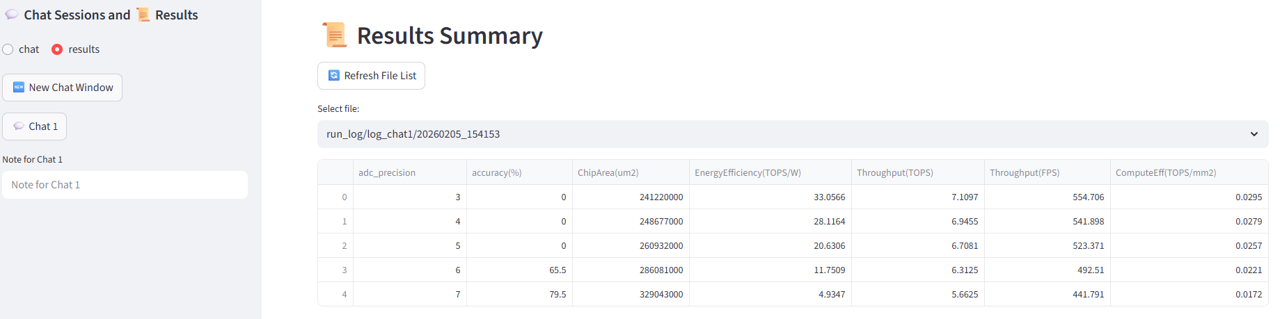}
  \caption{Dialogue example: simulation results.}
  \label{appendixDialogue4}
\end{figure}

This Appendix section shows the ChatNeuroSim User Interface (UI) and a dialogue example. The UI overview of the sidebar, chat area, and input bar is shown in Fig. \ref{appendixDialogue1}. In the example, the user inputs a request of Testbench Auto-design to check the CIM PPA under different ADC precisions. The agent checks the missing and invalid parameters sends the feedback to the user (Fig. \ref{appendixDialogue1} and Fig. \ref{appendixDialogue2}). After the adjustment are provided by the user, the agent changes the current parameters and rechecks the missing and invalid parameters (Fig. \ref{appendixDialogue2}). The simulation is executed successfully after user's final confirmation (Fig. \ref{appendixDialogue3}). For the simulation results, the user can switch to the "results" toggle and find all configurations and corresponding PPA summary (Fig. \ref{appendixDialogue4}).

\end{document}